\documentclass[journal]{IEEEtran}
\usepackage{lineno}
\modulolinenumbers[1]
\usepackage{graphicx}
\usepackage{url}
\usepackage{color} 
\usepackage[table]{xcolor}
\usepackage{afterpage}  
\usepackage{longtable,pdflscape,booktabs}
\usepackage{epstopdf}
\usepackage{xcolor}
\usepackage{hhline}
\usepackage{float} 
\usepackage{amsmath} 
\usepackage{wrapfig}
\usepackage{multirow} 
\usepackage{caption}
\usepackage{subcaption} 
\usepackage{tabularx}
\usepackage{booktabs}
\usepackage{multicol}
\usepackage{booktabs}
\usepackage{array}
\usepackage[dvipsnames]{xcolor}
\usepackage{soul}
\usepackage{comment}
\usepackage{xltabular}
\usepackage{cleveref}
\usepackage{float}
\usepackage{cite}
\usepackage{array}
\usepackage{lettrine}
\usepackage{amsthm}
\usepackage{booktabs} 
\usepackage{balance}
\usepackage{optidef}
\usepackage{lineno} 
\usepackage{amssymb}

\usepackage{flushend}
\usepackage{xspace}
\usepackage{enumitem}

\usepackage[most]{tcolorbox} 
\usepackage{framed}
\usepackage[normalem]{ulem}

\usepackage{xr}

\newcommand{\noop}[1]{}

\newif\ifshowmods

\showmodstrue     

\ifshowmods

	\newcommand{\textstrike}[1]{\st{#1}}
\else

	\newcommand{\textstrike}[1]{}
\fi

\newlist{todolist}{itemize}{2}
\setlist[todolist]{label=$\square$}
\usepackage{booktabs}

\begin{document}

\title{Energy-Efficient Code Generation Using Large Language Models: A Systematic Literature Review}

\author{Sabiya Banu Masthan Ali, Syed Muhammad Danish, Aroosa Hameed, Oussema Kirmani, Gautam Srivastava
\IEEEcompsocitemizethanks{
\IEEEcompsocthanksitem Sabiya Banu Masthan Ali, Oussema Kirmani, Aroosa Hameed and Syed Muhammad Danish are with Algoma University, Brampton, Canada. (email: \{smasthanali, okirmani, aroosa.hameed, syed.danish\}@algomau.ca)
\IEEEcompsocthanksitem Gautam Srivastava is with Department of Math and Computer Science, Brandon University, Brandon, Canada (email: srivastavag@brandonu.ca).

}}




\maketitle

\begin{abstract}
Large Language Models (LLMs) are increasingly used in software engineering for code generation, completion, translation, and repair. While substantial research has examined the energy consumption and environmental impact of LLM training and inference, considerably less attention has been given to the energy efficiency of the code produced by these models. Existing studies are fragmented across different models, programming tasks, benchmarks, hardware platforms, and measurement methods, making it difficult to draw consistent conclusions about the energy efficiency of LLM-generated code. No systematic synthesis currently consolidates these findings or identifies common trends, gaps, and methodological inconsistencies. To address this gap, we conduct a systematic literature review of existing empirical studies on the energy efficiency of LLM-generated code. We examine the extent to which LLMs generate energy-efficient code, the metrics used for evaluation, the benchmarks, datasets, and measurement tools employed, and the influence of prompting and fine-tuning strategies. Our findings show that the current evidence remains limited and highly heterogeneous. Most studies rely on conventional software-efficiency metrics and existing programming or code-generation benchmarks rather than evaluation methods specifically designed for LLM-generated code. The reported energy-efficiency outcomes are strongly dependent on the evaluated model, task, benchmark, and experimental configuration, and no single prompting strategy consistently improves energy efficiency across settings. We further find limited investigation of small language models, energy-aware fine-tuning, diverse programming languages, real-world software domains, and direct hardware-based energy measurement. Overall, the review highlights the lack of standardized evaluation practices and dedicated energy-efficiency benchmarks for LLM-generated code and identifies the need for more reproducible, hardware-aware, and context-sensitive evaluation methods.

\end{abstract}



\section{Introduction}

Large Language Models (LLMs) are advanced AI systems trained on large corpora of data using transformer-based architectures \cite{vaswani2017attention, hameed2025block}. These models are capable of understanding, reasoning over, and generating human-like language, enabling their use across a wide range of applications\cite{zhao2023llmsurvey}. In recent years, LLMs have gained significant attention in the software engineering domain, where they support and automate various programming tasks. One of the most prominent applications is code generation, in which an LLM produces executable source code from natural language instructions \cite{jiang2022discovering}. Closely related tasks include code completion (predicting missing or subsequent lines of code), code translation (converting source code between programming languages), bug fixing (detecting and correcting programming errors), and documentation generation (producing natural language explanations of code functionality). These capabilities are now embedded in widely used AI-assisted development tools, such as GitHub Copilot \cite{github_copilot_docs}, CodeGeeX \cite{zai_codegeex_repo}, and Amazon CodeWhisperer \cite{aws_codewhisperer_docs}, which leverage state-of-the-art code LLMs to improve developer productivity and streamline the software development process.

LLMs \cite{otoum2025llms, Hameed2025UAVLLM} have achieved tremendous success in code generation \cite{herrington2003code, rahman2025refactorcoderqa}; however, there is growing concern about the environmental impact of their development and use. Training and deploying LLMs can incur substantial energy consumption, carbon emissions, and water usage \cite{desislavov2023trends, morrison2503holistically}. Once deployed, these models continue to consume energy during inference, and their overall environmental impact increases with usage frequency \cite{desislavov2023trends}. Although LLM inference itself consumes considerable energy, the environmental impact does not end at the inference stage. Once integrated into software systems, the efficiency of the generated code becomes an important factor in long-term operational energy consumption \cite{mehra2023assessing}. Inefficient code can increase computational load, memory usage, and execution time, thereby increasing energy consumption during execution. This is particularly relevant given the growing contribution of the ICT sector to global greenhouse gas emissions \cite{freitag2021real}. The growing adoption of AI-assisted code generation further increases the importance of evaluating the energy efficiency of generated code \cite{daniotti2025using}. As LLM-generated code becomes increasingly integrated into software development workflows, it is important to determine whether these models can generate code that is not only functionally correct but also energy efficient during execution.

While a growing number of studies have examined the energy efficiency and environmental impact of LLM-generated code, the literature still lacks a detailed and systematic review of this area. To address this gap, we present a systematic literature review (SLR) that identifies, collects, and organizes research on the energy efficiency of code generated by LLMs. Specifically, we review four types of contributions. First, we investigate studies evaluating whether LLMs can generate code with lower energy consumption and resource usage. Second, we examine the metrics and parameters used to evaluate generated code, including energy consumption, power consumption, and carbon emissions, etc. Third, we identify the benchmarks, datasets, and measurement tools used for energy-related evaluation of LLM-generated code. Finally, we examine how prompting strategies and fine-tuning methods influence the energy and resource efficiency of generated code. In this SLR, we specifically focus on energy efficiency as the primary dimension of environmental and operational sustainability. Broader software quality attributes, such as maintainability and evolvability, are outside the primary scope of this review unless they are directly evaluated in relation to energy or resource efficiency.

To collect and review the relevant literature, we follow a systematic and structured process. We begin by identifying an initial set of studies through database searches and reference checking. We then apply clear inclusion and exclusion criteria in multiple screening stages to select the most relevant papers. For the selected studies, we extract data using predefined forms to ensure consistency and accuracy. We organize the collected information into tables and visual summaries to compare key characteristics of the reviewed studies. Based on this analysis, we identify common patterns, main findings, and existing limitations in the current body of research.

Overall, we find that research in this domain remains limited. The diversity of evaluated models is narrow, with most studies focusing on LLMs, while small language models receive much less attention. Energy measurements are primarily conducted at the software level, and only a few studies use external hardware tools for a direct physical measurement. In addition, most research focuses on a small set of programming languages and general-purpose software development, with limited attention given to other software domains. We also observe the absence of dedicated energy-focused benchmarks designed specifically to evaluate the energy efficiency of LLM-generated code. Finally, energy-aware fine-tuning remains largely unexplored and represents an important direction for future research. By synthesizing the current body of research, we provide a clear overview of the models studied, the programming languages considered, the evaluation methods applied, and the limitations that remain. This structured analysis helps identify research gaps and offers direction for future work toward more sustainable and environmentally responsible code generation using LLMs.

The remainder of this paper is organized as follows. Section II provides the related work. Section III describes the research methodology. Section IV presents the results from the extracted data. Section V provides insights into the extracted data. Section VI presents the future research directions. Section VII outlines the validity threats and limitations. Finally, Section VIII presents concluding remarks.

\section{Related Work}
\label{sec:rw}

\subsection{LLMs for Code Generation}
Code generation refers to the automatic production of executable source code from inputs such as natural language descriptions, design documents, or existing code, with the goal of reducing manual effort and improving development efficiency. Traditional approaches based on rule-based systems, templates, or domain-specific languages provide partial automation for well-defined tasks but have limited contextual understanding, struggle to generate complete and logically consistent code, and lack adaptability to diverse development needs \cite{becker2023programming, xu2022ide, huynh2025large}. These limitations reduce their effectiveness for modern software development, which increasingly involves open-ended requirements, evolving APIs, and complex architectures \cite{dong2025survey}.

LLMs have recently emerged as a powerful approach for code generation, supporting source-code generation from natural language or existing code across diverse programming languages and development environments. Their applications include code completion \cite{bruch2009learning}, code optimization \cite{ishibashi2024self}, code refactoring \cite{tapader2025code}, code translation \cite{luo2024bridging}, automated program repair \cite{bouzenia2025repairagent}, test generation \cite{ryan2024code}, debugging \cite{tian2024debugbench}, and code explanation \cite{nam2024using}. Research on LLM-based code generation has also examined model adaptation \cite{ma2026llamoco}, prompting strategies \cite{li2025structured}, correctness and code quality \cite{santa2025llm}, and developer productivity \cite{weisz2025examining}. Overall, LLMs are becoming increasingly important programming assistants, with growing attention to the correctness, efficiency, and practical use of generated code.

\subsection{Energy-Efficient and Green Software Engineering}
software engineering is influenced not only by hardware but also by software design and implementation decisions. Prior research has therefore examined energy efficiency across the software development lifecycle, including requirements, design, implementation, verification, and maintenance \cite{georgiou2019software}.

Energy-efficient software development involves identifying energy-intensive code and applying suitable optimization techniques. Existing approaches use profiling \cite{7965243}, source-code analysis \cite{9826760}, and energy measurement tools to locate energy hot spots \cite{noureddine2015monitoring} and guide changes to algorithms \cite{manotas2014seeds}, and compiler configurations \cite{abdulsalam2014program}. Related work has also investigated energy-aware refactoring \cite{10.1145/2652524.2652538}, design patterns \cite{6224257}, antipatterns \cite{morales2016anti}, and code smells \cite{palomba2019impact} to identify recurring practices associated with unnecessary computation, memory use, I/O, or inefficient resource management.

Energy efficiency is commonly evaluated using direct measures such as energy consumption and power, together with conventional software-efficiency metrics such as execution time, CPU utilization, memory usage, and hardware performance counters \cite{kern2013green}. However, conventional performance metrics do not directly measure energy consumption. In particular, lower execution time does not necessarily imply lower energy use, as energy consumption may not be directly correlated with CPU time \cite{10.1007/978-3-030-45237-7_8}. Therefore, direct energy measurement provides stronger evidence when determining whether a software implementation is genuinely more energy efficient.

These findings provide an important foundation for studying LLM-generated code. Green software engineering has already identified several software-level factors that can influence energy consumption, including algorithmic choices, data structures, parallelism, programming-language characteristics, coding practices, and refactoring. An important question is therefore whether these established principles are considered when LLMs generate or optimize code, and whether their effects are evaluated using direct energy measurements rather than performance indicators alone.

\subsection{Energy and Performance Efficiency of LLM-Generated Code}
Recently, several studies have investigated the energy and performance efficiency of LLM-generated code. Cheung et al.~\cite{cheung2025comparative} compared the environmental impact of LLM-generated and human-written code and found that generated code can require greater computational resources and energy. Islam et al.~\cite{islam2025evaluating} evaluated 20 LLMs across 878 LeetCode problems and showed that functionally correct generated code can still be less energy efficient than human-written solutions. Tuttle et al.~\cite{tuttle2024can} similarly compared eight LLMs across different LeetCode problems and prompting strategies.

Other studies investigate techniques for improving code efficiency. Cappendijk et al.~\cite{cappendijk2025exploration} examined how prompt-engineering strategies affect the energy consumption of LLM-generated Python code and found that certain prompt modifications can reduce energy use. Peng et al.~\cite{peng2024large} used LLMs to refactor existing code for improved energy efficiency while preserving correctness and performance, while Rani et al.~\cite{rani2025can} evaluated several LLMs for optimizing real-world MATLAB programs in terms of energy consumption, memory usage, execution time, and correctness.

Researchers have also proposed metrics for comparing generated code with reference implementations. For example, Tuttle et al.~\cite{tuttle2024can} introduced \textit{RuntimeRatio} and \textit{EnergyRatio} to quantify relative runtime and energy consumption. Overall, existing research remains fragmented across models, tasks, metrics, benchmarks, and measurement approaches. This motivates a systematic synthesis that distinguishes direct energy evaluation from conventional performance measures and examines how LLM-based approaches evaluate and improve the energy efficiency of generated code.

\subsection{Existing Reviews and Positioning of This SLR}
To position this SLR within the existing literature, we examined prior systematic reviews and surveys related to energy-efficient LLM-based code generation and software engineering. We identified several recent reviews covering related areas, including general LLM-based code generation \cite{danyaro2025llm,bacin2025systematic,bistarelli2025usage,alharbi2026automatic,jiang2024llm_code_generation_survey,zan-etal-2023-large,chen2024survey_eval_llm_codegen,jabrw2025systematic_survey_llm_code_generation}, LLMs in software engineering \cite{hou2024large,fan2023llm4se_survey_open_problems}, code completion \cite{husein2025large}, automated program repair \cite{zhang2024systematic,yang2025llm_apr_survey,zhang2024llm_apr_slr}, parameter-efficient fine-tuning for large code models \cite{afrin2025systematic}, domain-specific and low-resource code generation \cite{joel2024survey}, LLMs in programming education \cite{pereira2025systematic,cambaz2024ai_programming_education,llm_cseducation_2025_slr,agbo2025computing_education_genai}, developer productivity \cite{mohamed2025impact,peng2023ai_developer_productivity}, and benchmarks for LLM-generated code \cite{paul2024benchmarks_metrics_codegen}. In addition, systematic mapping studies have examined the use of LLMs for generating, reviewing, and translating code \cite{albuquerque2024generating,saei2025automated_codegen_sms,codereview_benchmarks_2026,rgaguena2025code_translation_sms}.

\noindent\textbf{Research Gap and Novelty of This SLR: }Despite the growing number of secondary studies on LLM-based code generation, we did not identify any SLR that specifically examines the energy efficiency of LLM-generated code. Existing reviews mainly focus on code-generation capability, correctness, software quality, benchmarks, developer productivity, or broader applications of LLMs in software engineering. As a result, they do not systematically synthesize whether LLMs generate energy-efficient code, how this energy efficiency is evaluated, which metrics, benchmarks, datasets, and tools are used, or whether these evaluation approaches are specifically designed for the LLM-generation context. They also do not provide a focused synthesis of how prompting and fine-tuning influence the energy efficiency of generated code. To the best of our knowledge, this study is the first SLR to examine these aspects together and provide a systematic view of the current evidence, evaluation practices, and optimization approaches for energy-efficient LLM-generated code. By doing so, our work addresses an important gap in the literature, as inefficient LLM-generated code can increase computational overhead and energy consumption during execution. Understanding and improving the energy efficiency of generated code is therefore important for supporting more energy-aware and environmentally responsible AI-assisted software development.

\section{Research Methodology}
\subsection{Overview}

We design our own SLR following the guidelines proposed by Kitchenham \cite{kitchenham2002preliminary}. In the first stage, we defined a review protocol, which included: (a) research questions, (b) a search strategy with a search string and relevant paper repositories, (c) inclusion and exclusion criteria, (d) an iterative study selection process, and (e) a data extraction strategy for classifying the primary studies. The process went through several cycles of execution, evaluation, and refinement until we finalized the protocol, which is described in the remainder of this section.



\begin{table}[t!]
\centering
\caption{Research Questions}
\label{tab:RQs}
\begin{tabular}{p{0.6cm}p{7.4cm}}
\hline
\textbf{RQ} & \textbf{Description} \\
\hline
\hline
RQ1 & To what extent do LLMs generate energy-efficient code? \\
RQ2 & What metrics are used to evaluate the energy efficiency of LLM-generated code, and to what extent are they specific to the LLM-generation context?  \\
RQ3 & What benchmarks, datasets, and tools are used to evaluate the energy efficiency of LLM-generated code, and which of them are specifically designed for energy-efficiency evaluation?  \\
RQ4 & How do prompting strategies or fine-tuning methods influence the energy-efficiency of the generated code? \\
\hline
\end{tabular}
\end{table}

\subsection{Research Questions}
To guide the review, we first define an overarching research question that captures the main objective of this SLR: \textit{What does the existing literature reveal about the energy efficiency of LLM-generated code and the methods used to evaluate and improve it?} This main question is further decomposed into four research questions that examine the effectiveness of LLMs in generating energy-efficient code, the metrics used for evaluation, the benchmarks and measurement tools employed, and the influence of prompting and fine-tuning strategies. The following four research questions (RQs) were defined to guide the SLR in Table \ref{tab:RQs}.

\begin{figure}[t!]
    \centering
    \includegraphics[width=\linewidth]{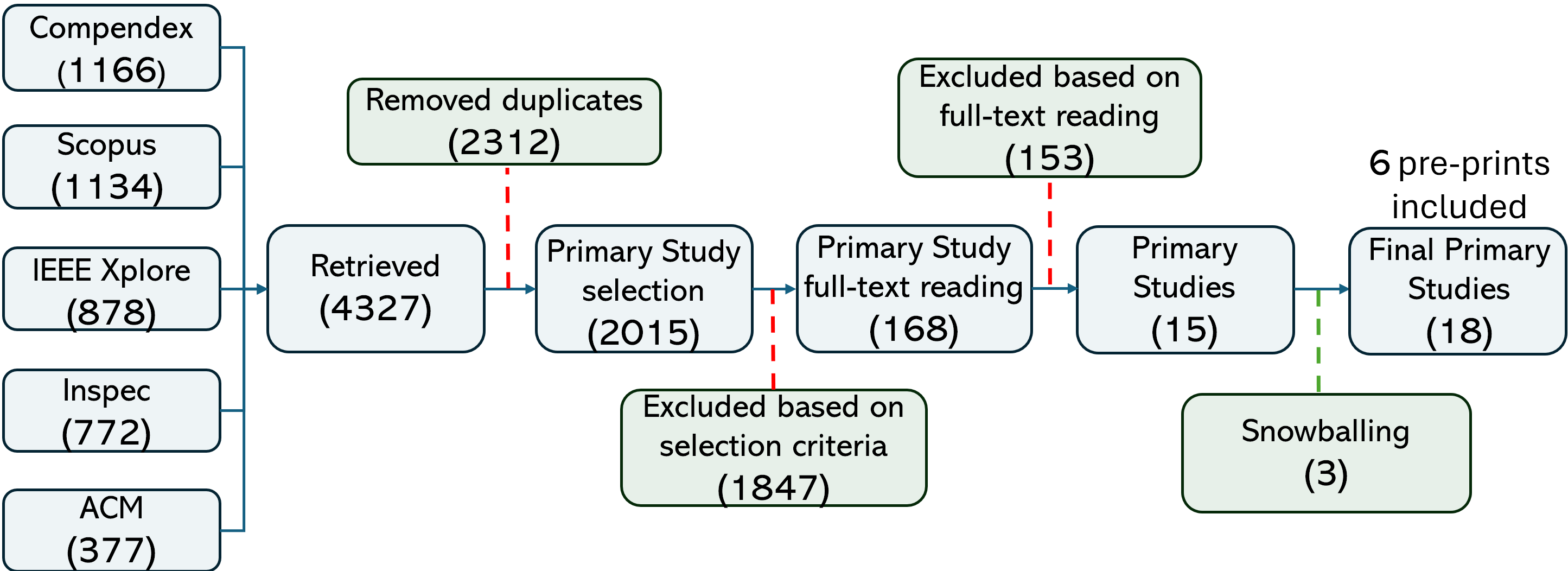}
    \caption{Primary Study Selection (red indicates removals and green indicates additions)}
    \label{fig:protocol}
\end{figure}

\textit{RQ1} examines the extent to which LLMs generate energy-efficient code based on empirical evidence reported in existing studies. It considers energy consumption and related operational indicators, such as power consumption, and carbon emissions, etc., when reported. It also examines how these outcomes vary across models, programming tasks, benchmarks, and programming languages.

\textit{RQ2} identifies the metrics used to evaluate the energy efficiency of LLM-generated code and examines whether these metrics are specific to the LLM-generation context or are conventional software-efficiency measures applied to generated code. In particular, we examine energy-related metrics and their relationship with performance, resource usage, and functional correctness, as well as evaluation measures that explicitly incorporate characteristics of generated outputs, such as comparison with reference solutions or multiple candidate generations.

\textit{RQ3} investigates the benchmarks, datasets, and measurement tools used to evaluate the energy efficiency of LLM-generated code. It examines whether these resources were specifically designed for energy-related evaluation of generated code or whether existing programming, code-generation, and performance benchmarks were adapted by incorporating energy measurements.

\textit{RQ4} examines the impact of model adaptation techniques, such as prompt engineering and fine-tuning, on the sustainability of generated code. It aims to identify practices that improve efficiency while maintaining correctness.

\subsection{Search Strategy}
Guided by the RQs, we adopt an iterative search strategy that queries digital libraries using boolean combinations of relevant keywords. The keywords are refined after each iteration to improve coverage, and the process continues until all manually identified publications are retrieved.
\begin{table}[t!]
\centering
\captionsetup{font=small}
\caption{Search String}
\label{table:searchstring}

\begin{tabular}{p{0.95\columnwidth}}
\hline
\scriptsize

("large language model*" OR "LLM" OR "Small Language Model*" OR "SLM" OR "GPT" OR "Codex" OR "ChatGPT" OR "code generation" OR "code generation model*" OR "AI code assistant" OR "AI-assisted programming" OR "automated programming" OR "code refactoring" OR "code optimization" OR "code synthesis" OR "program synthesis" OR "code completion" OR "software language model" OR "AI programming tool*" OR "generative AI for code" OR "code LLM" OR "AI-based code generation" OR "prompt engineering" OR "instruction tuning" OR "fine-tuning" OR "zero-shot learning" OR "few-shot prompting" OR "chain of thought" OR "model adaptation" OR "Llm-generated code")

\par\smallskip
\textbf{AND}
\par\smallskip

("sustainable code" OR "green software" OR "green software engineering" OR "energy efficient" OR "eco-friendly software" OR "low-energy code" OR "energy-aware" OR "carbon footprint" OR "environmental impact" OR "code sustainability" OR "energy-optimized software" OR "energy-efficient software" OR "sustainability in software engineering" OR "sustainable programming" OR "energy efficiency" OR "efficient code")

\\
\hline
\end{tabular}

\end{table}

\subsubsection{Search String}
Our search string covered two main aspects: (i) LLMs, code generation, and related techniques, and (ii) sustainability and energy efficiency. We validated the search using a benchmark set of 13 manually identified studies \cite{peng2024large, rubei2025prompt, tuttle2024can, dearing2025leveraging, cappendijk2025exploration, coignion2024green, cheung2025comparative, rani2025can, podder2025empirical, peng2025sysllmatic, stivala2024investigating, ashraf2025energy, ashraf2025toward}. Of these 13 studies, two \cite{ashraf2025energy, ashraf2025toward} were our prior works in this research area. While conducting these studies, we identified the remaining 11 studies through our review of the related literature. Together, these 13 studies formed the benchmark set used to validate the search strategy. The initial query retrieved 10 of the 13 studies. To capture the missing studies, additional terms (e.g., “LLM-generated code,” “energy efficiency,” and “efficient code”) were incorporated. The revised search string successfully retrieved the complete benchmark set and was therefore adopted for the remainder of the study. The final search string is presented in Table~\ref{table:searchstring}.

We searched five databases, IEEE Xplore, Compendex, Inspec, Scopus, and the ACM Digital Library, and managed the retrieved publications using an Excel spreadsheet. The generic search expression was adapted to the search syntax and interface of each digital library while preserving the same conceptual search domains. In Engineering Village, the same search query was used and the results were filtered separately for Compendex and Inspec, as reported in Table~\ref{table:searchstring}. For Scopus, the search expression was adapted using the TITLE-ABS-KEY field to search titles, abstracts, and keywords. In the ACM Digital Library, the two main parts of the query were entered separately using the Search Within fields; by default, ACM applies an AND relationship between multiple Search Within terms. For IEEE Xplore, the expression was adapted to the limitations of Command Search, including restrictions on the number of terms allowed per clause and wildcard usage. The exact database-specific search expressions are provided in the supplementary material to facilitate replication. The search process and publication counts are shown in Fig.~\ref{fig:protocol}.

\subsection{Inclusion and Exclusion Criteria}

We applied predefined inclusion and exclusion criteria, summarized in Table~\ref{tab:inc_exc}, to identify studies relevant to the energy efficiency of LLM-generated code. The criteria were applied during initial screening, full-text review, and snowballing. To maintain a consistent empirical scope, studies were required to include an energy-related evaluation of generated code. Studies addressing software efficiency or quality without empirical energy-related analysis were therefore excluded. Relevant arXiv preprints were also considered because this research area is still emerging and may not yet be fully represented in peer-reviewed venues.
\begin{table}[htbp]
\centering
\caption{Inclusion and Exclusion Criteria}
\label{tab:inc_exc}

\begin{tabularx}{\columnwidth}{X}
\toprule
\textbf{Inclusion Criteria} \\
\midrule
\textbf{--} The study explicitly uses, evaluates, or fine-tunes an LLM for code-related tasks. \\
\textbf{--} The study addresses energy-related aspects related to code generation. \\
\textbf{--} The study includes empirical evaluation of generated code. \\
\textbf{--} The full text is accessible. \\
\midrule
\textbf{Exclusion Criteria} \\
\midrule
\textbf{--} The study does not involve an LLM or focuses only on traditional ML/DL/NLP approaches. \\
\textbf{--} The study does not assess energy-related aspects. \\
\textbf{--} The study focuses on tasks unrelated to code generation (e.g., text summarization, image captioning, question answering). \\
\textbf{--} The study discusses software efficiency or quality but does not perform an empirical energy-related evaluation of the generated code. \\
\textbf{--} The study addresses sustainability in domains unrelated to code generation (e.g., green supply chain management). \\
\textbf{--} The study is purely theoretical or conceptual without empirical validation. \\
\textbf{--} The publication is a doctoral symposium paper, editorial, position paper, or poster. \\
\textbf{--} The study is a survey, review, or tutorial without original experimental results. \\
\textbf{--} The full text is not accessible or is not written in English. \\
\bottomrule
\end{tabularx}
\end{table}



\begin{table*}[!ht]
\centering
\footnotesize
\caption{Data Extraction Form}
\label{tab:dataform}
\begin{tabular}{llp{10cm}}
\toprule
\textbf{Category} & \textbf{Data Items} & \textbf{Description} \\
\midrule
\multirow{3}{*}{General} 
 & Context of Study 
 & Categorized as an academic or industrial setting. \\
 
 & Code-Related Task 
 & The primary coding task addressed (e.g., code generation, completion, etc.)\\
 
 & Coding Application 
 & The application domain of generated code (e.g., blockchain, scientific computing, etc). \\
\midrule

\multirow{3}{*}{Model}  
 & Evaluated Models and Size Categories 
 & Lists and classify the LLM(s) used in the study, along with their parameter sizes. \\  
 
 & Model Openness 
 & Specifies whether the models are open or closed-source, or a combination of both. \\  
 
 & Model Domain Specialization 
 & Indicates whether the model is general-purpose, code-specific, or a combination of both. \\  
\midrule
 
\multirow{7}{*}{Evaluation} 
 & Evaluation Metrics 
 & Metrics used to assess performance and energy-efficiency. \\
 
 & Custom Evaluation Metric 
 & Indicates whether the study introduces a novel evaluation metric. \\
 
 & Libraries / Tools for Measurement 
 & Tools or libraries used for metric collection. \\
 
 & Hardware Configuration 
 & Description of the hardware and experimental setup used for code execution. \\
 
 & Experimental Configuration 
 & Details regarding experiment repetitions and inference parameters (e.g., temperature, top-p, maximum tokens, batch size, precision). \\
 
 & Open-Source Availability  
 & Availability of code, prompts, datasets, or other resources to support reproducibility. \\
 
 & Benchmark Usage  
 & Indicates whether established benchmarks are used for model evaluation. \\
\midrule

\multirow{2}{*}{Model Adaptation}  
 & Fine-Tuning Details 
 & Indicates whether the study is performing energy-aware fine-tuning. \\
 
 & Use of Prompt Engineering 
 & Whether prompt engineering strategies are applied to generate energy-efficient code. \\
\bottomrule
\end{tabular}
\end{table*}

\subsection{Study Selection Procedure}
After applying the search string (Table~\ref{table:searchstring}), we initially retrieved 4,327 publications. Removing duplicates reduced this number to 2,015. This refined set was then screened by title and abstract to identify the initial pool of primary studies. To improve the reliability of this manual filtering process, we iteratively refined the inclusion and exclusion criteria. First, we randomly selected a sample of 100 publications from the 2,015 retrieved studies. Three authors independently reviewed these papers against the criteria and decided whether they should be included in the initial pool. We then measured inter-rater agreement using Fleiss’ Kappa \cite{fleiss1981measurement}. If the agreement was low, the authors discussed the disagreements, revised the criteria, and repeated the process until an acceptable level of inter-rater reliability was achieved.

Two such rounds emerged. Our initial Fleiss’ Kappa index was 0.378, indicating low agreement. Post-analysis showed that the main issues were unclear selection criteria and inconsistent interpretations of sustainability-related concepts. Disagreements arose from including papers on energy efficiency in domains other than code generation, as well as studies on DNNs or CNNs, which were outside our scope. Another challenge was the definition of sustainability, where some studies focused solely on efficiency without explicitly addressing sustainability. We refined the inclusion criteria to focus strictly on LLMs and code generation by excluding studies on general neural networks and other application domains, and then conducted a second iteration. In the second iteration, Fleiss’ Kappa inter-rater agreement increased to 0.833, indicating that the researchers had reached a sufficient and consistent understanding of the inclusion criteria. The remaining publications were then evenly distributed between the two researchers, with each researcher screening their assigned studies using the refined criteria. This process resulted in 168 studies being selected for full-text assessment.

Full-text screening was then performed on the 168 studies selected during the title and abstract screening stage. To ensure consistency, the same pilot process with Fleiss’ Kappa measurement was applied at this stage. A random sample of 30 studies was independently assessed by two reviewers, yielding a Kappa value of 0.70, indicating substantial agreement. Disagreements and uncertain cases were discussed between the reviewers until consensus was reached and a common understanding of the eligibility criteria was established. The remaining studies were then evenly distributed between the two reviewers, who performed full-text assessment on their assigned studies using the agreed criteria. Following full-text screening, 15 studies satisfied the eligibility criteria and were retained as primary studies.

\paragraph{Snowballing}

We conducted a snowballing search \cite{jalali2012systematic} to identify additional studies that may not have been retrieved through the database search. Snowballing was performed after the initial screening process had identified 15 primary studies. Both backward and forward snowballing were applied to each of these studies. For backward snowballing, we examined the reference lists of the selected studies to identify earlier publications that were potentially relevant to the scope of this review. For forward snowballing, we used Google Scholar to identify subsequent publications that cited each of the selected studies. All potentially relevant references and citing publications identified through this process were assessed using the same predefined inclusion and exclusion criteria applied during the main study-selection process. Candidate studies were first examined based on their title and abstract, followed by full-text assessment when their relevance could not be determined from the initial screening. Studies were included only if they satisfied the same eligibility requirements as studies identified through the database search, particularly the requirement for an empirical energy-related evaluation of LLM-generated code.

The snowballing process identified three additional eligible studies, increasing the final set from 15 to 18 primary studies. Snowballing was conducted by one reviewer and was used as a complementary search strategy rather than as the primary study-identification mechanism. Its purpose was to reduce the possibility of missing relevant studies due to differences in indexing or terminology. Fig.~\ref{fig:protocol} summarizes the overall study-selection process, from the initial database search to the final set of 18 primary studies used for data extraction and synthesis.


\begin{table*}[ht]
\centering
\caption{Quality Assessment Criteria and Summary of Assessment Results}
\label{tab:quality_assessment}
\begin{tabularx}{\textwidth}{l X c c c}
\toprule
\textbf{ID} & \textbf{Quality Assessment Question}
& \textbf{Yes} & \textbf{Partially} & \textbf{No} \\
\midrule

QA1 & Are the study objectives and research questions clearly defined? 
& 18 (100\%) & 0 (0\%) & 0 (0\%) \\

QA2 & Are the evaluated LLMs, code-generation tasks, datasets or benchmarks, and experimental settings sufficiently described?  
& 18 (100\%) & 0 (0\%) & 0 (0\%) \\

QA3 & Does the study evaluate the functional correctness or validity of the generated or optimized code before comparing its energy efficiency? 
& 13 (72.2\%) & 2 (11.1\%) &   3 (16.7\%)  \\

QA4 & Is the hardware and execution environment sufficiently described?
& 15 (83.3\%) & 1 (5.6\%) & 2 (11.1\%) \\

QA5 & Are the experimental and generation settings sufficiently described?
& 6 (33.3\%) & 9 (50.0\%) & 3 (16.7\%) \\

QA6 & Are appropriate baselines and evaluation measures used?
& 17 (94.4\%) & 1 (5.6\%) & 0 (0\%) \\

QA7 & Is the experimental design and analysis sufficient to support the reported findings?
& 12 (66.7\%) & 6 (33.3\%) & 0 (0\%) \\

QA8 & Are sufficient artifacts provided to support reproducibility?
& 11 (61.1\%) & 0 (0\%) & 7 (38.9\%) \\

QA9 & Are the prompting, generation, fine-tuning, or other model-adaptation settings sufficiently described when applicable?
& 16 (88.9\%)   & 0 (0\%)  & 2 (11.1\%) \\

QA10 & Are limitations, threats to validity, or factors affecting the interpretation and generalizability of the results discussed?
& 13 (72.2\%) & 5 (27.8\%)  & 0 (0\%) \\

\bottomrule
\end{tabularx}
\end{table*}

\subsection{Data Extraction}

We created a structured data extraction form to collect information from each primary study consistently and accurately, focusing on our research questions (RQ1--RQ4). First, three researchers independently prepared draft extraction forms based on a small sample of studies. The researchers then discussed and combined these drafts to develop the final extraction form shown in Table~\ref{tab:dataform}. The categories in the form were designed to capture the main characteristics and differences identified in research on sustainable code generation using LLMs.

Similar to the study selection process, a pilot extraction was conducted before extracting data from all primary studies. The researchers independently extracted information from a common sample of five studies using the initial extraction form. The extracted results were compared through joint discussions to identify differences in interpretation, inconsistencies in categorization, and ambiguous fields. Based on these discussions, the extraction guidelines were refined to establish a consistent interpretation of each field.

Given the relatively small number of primary studies (18), one researcher performed the complete data extraction for all studies, while a second researcher independently cross-checked all extracted values against the original studies. Any discrepancies, ambiguous cases, or unclear categorizations identified during extraction or cross-checking were jointly discussed and resolved by consensus. Only information explicitly reported in the primary studies was recorded; unavailable information was marked as \textit{Not Specified} rather than inferred. The finalized and cross-checked dataset was then used for data synthesis and analysis across the research questions.

\subsection{Data Extraction Form}

To improve consistency and reproducibility, we used predefined rules to categorize the characteristics of the selected studies during data extraction. The categories were assigned based on information reported in the primary studies and included study context, code-related task, application domain, model size, model availability, model specialization, evaluation hardware, and experimental configuration. A study could be assigned to more than one category when applicable. If the information reported in a study was insufficient for classification, it was marked as ``Not Specified.'' These categories were used to organize and summarize the literature and were not intended to establish causal relationships with energy efficiency.


The extracted data offers a clear framework for analysis, enabling the comparison and summarization of findings across all studies. The results are presented in the next section.

\subsection{Quality Assessment of Primary Studies}

To assess the methodological quality of the selected primary studies, we conducted a quality assessment using ten criteria (QA1--QA10). Each criterion was evaluated as "Yes", "Partially", or "No". A "Partial" rating was assigned when a study addressed the criterion but provided incomplete evidence, reporting, or coverage. The discussion below focuses on criteria for which at least one study received a Partial rating; the complete aggregated results are reported in Table~\ref{tab:quality_assessment}.

QA3 examined whether the functional correctness of the generated or optimized code was verified before comparing its energy efficiency. Two studies (11.1\%) were rated Partial. \cite{rubei2025prompt} evaluated generated code using Exact Match and Edit Distance against ground-truth completions but did not perform compilation- or execution-based functional testing. Similarly, \cite{cappendijk2025exploration} provided some evidence of code validity but did not establish comprehensive functional correctness before energy-efficiency evaluation. QA4 considered the completeness of the reported hardware and execution environments. One study (5.6\%), \cite{cheung2025comparative}, was rated Partial because it reported some hardware assumptions, such as a standard laptop with 16 GB RAM, but did not fully specify the execution environment or GPT-4 backend hardware. Its energy consumption values were also derived from estimates rather than direct measurements. QA5 assessed whether the experimental and generation settings were sufficiently reported. Nine studies (50.0\%) were rated Partial. These studies generally reported the prompting procedure, number of runs, or generation attempts but omitted important decoding parameters such as temperature, top-p, top-k, or token limits \cite{rubei2025prompt, cheung2025comparative, peng2024large}.

QA6 examined whether appropriate baselines and evaluation measures were used. One study (5.6\%), \cite{vartziotis2024carbon} was rated Partial because, although relevant evaluation measures were used, it lacked a strong comparative baseline and relied partly on estimated rather than directly measured LLM energy consumption. QA7 examined whether the experimental design and analysis provided sufficient evidence to support the reported findings. Six studies (33.3\%) were rated Partial, primarily due to limited benchmark coverage, small numbers of programming tasks, or restricted model diversity. Some studies evaluated only three to six programming problems \cite{cappendijk2025exploration, peng2024large, cursaru2024controlled}, while others relied on a single LLM or code assistant \cite{cursaru2024controlled, vartziotis2024carbon, podder2025empirical, cheung2025comparative}. Finally, QA10 considered the completeness of reported limitations, threats to validity, and factors affecting interpretation and generalizability. Five studies (27.8\%) were rated Partial because limitations were acknowledged but not discussed systematically (\cite{tuttle2024can,dearing2025leveraging, vartziotis2024carbon, vartziotis2024learn, podder2025empirical}).

\section{Results}
\label{subsec:data_extraction}
We analyze data from 18 primary studies, including 12 peer-reviewed publications and 6 preprints. The extracted data, summarized in Table~\ref{tab:dataform}, support the synthesis and comparison of findings for \textit{RQ1--RQ4} in Section~\ref{subsec:results_rqs}.

\subsection{General}
\subsubsection{Context of Study}

We first categorized the selected studies based on the research environment, distinguishing between academic and industrial settings. A study was labelled as academic if all authors were affiliated with universities or research institutions, and as industrial if at least one author had an industry affiliation. The distribution is summarized in Table \ref{tab:context_grouped}.

\begin{table}[ht!]
    \centering
        \begin{tabular}{p{1.7cm} p{4.9cm} p{0.5cm}}
        \hline
        Category & Primary Studies & Total \\
        \hline
        \hline
        Academic & \cite{peng2024large, rubei2025prompt, tuttle2024can, dearing2025leveraging, cappendijk2025exploration, coignion2024green,
        cheung2025comparative, rani2025can, podder2025empirical, peng2025sysllmatic, stivala2024investigating, cursaru2024controlled,apsan2025generating}  & 13\\
        Industrial & \cite{vartziotis2024carbon, vartziotis2024learn, islam2025evaluating, ashraf2025energy, ashraf2025toward}  & 5 \\
        \hline
        \end{tabular}
        \captionsetup{font=small}
    \caption{Context of study}
    \label{tab:context_grouped}
\end{table}

\subsubsection{Code-Related Task}

We categorized the selected studies based on the code-related tasks they addressed. Code-related tasks refer to code generation, code completion, code optimization, and code refactoring. As several studies addressed more than one task (e.g., combining code generation with optimization or refactoring), a multi-label classification approach was adopted; therefore, a single study may appear in multiple task categories. Table~\ref{tab:code-related} shows the distribution of studies across code-related tasks. 

\begin{table}[ht!]
    \centering
        \begin{tabular}{p{2.5cm} p{5cm} p{0.5cm}}
        \hline
        Category & Primary Studies \\
        \hline
        \hline
        Code Generation &\cite{dearing2025leveraging, tuttle2024can, cappendijk2025exploration,coignion2024green,
        cheung2025comparative, vartziotis2024carbon, vartziotis2024learn, islam2025evaluating, podder2025empirical, ashraf2025energy, peng2025sysllmatic, stivala2024investigating, cursaru2024controlled, ashraf2025toward, apsan2025generating}

\\
        Code Optimization & \cite{peng2024large, rani2025can, podder2025empirical, stivala2024investigating}\\
        
        Code Completion & \cite{rubei2025prompt} \\
        Code Refactoring & \cite{dearing2025leveraging}  \\
        \hline
        \end{tabular}
        \captionsetup{font=small}
    \caption{Code-related tasks}
    \label{tab:code-related}
\end{table}

\subsubsection{Code Application Domain}

To better understand the application contexts of the selected studies, we categorized them based on the type of software and workloads they target. General-purpose software development includes programming tasks implemented using standard programming languages and execution environments, such as algorithmic problem solving and common application development, which are not tied to a specific scientific or engineering domain. In contrast, scientific and engineering applications focus on domain-specific workloads involving numerical computation, simulation, or parallel processing. The distribution of studies across these application contexts is presented in Table~\ref{tab:application}. 

\begin{table}[ht!]
    \centering
        \begin{tabular}{p{3cm} p{3.9cm} p{0.5cm}}
        \hline
        Category & Primary Studies & Total \\
        \hline
        \hline
        General-Purpose Software Development & \cite{peng2024large, rubei2025prompt, tuttle2024can, cappendijk2025exploration, coignion2024green, cheung2025comparative, vartziotis2024carbon, vartziotis2024learn, islam2025evaluating, podder2025empirical, ashraf2025energy,  peng2025sysllmatic, stivala2024investigating,  cursaru2024controlled, ashraf2025toward, apsan2025generating}  & 16\\
        Scientific and Engineering Applications & \cite{rani2025can}, \cite{dearing2025leveraging}  & 2 \\
        \hline
        \end{tabular}
        \captionsetup{font=small}
    \caption{Application domains targeted by the selected studies}
    \label{tab:application}
\end{table}

\subsection{Evaluation}

\subsubsection{Evaluation Metrics}
We categorize the evaluation metrics considered in the primary studies into two broader groups: energy-related metrics and other evaluation metrics. Energy-related metrics capture absolute and relative energy usage, power consumption, and environmental impact. Other evaluation metrics include performance, correctness, memory, code quality, and miscellaneous measures. Performance metrics \cite{worlton1991toward} measure execution time, latency, and throughput, while correctness metrics assess the functional validity of generated outputs. Memory metrics \cite{nguyen2020classification} quantify resource utilization, including memory usage and computational operations. Code quality metrics \cite{sharma2020we} evaluate lexical, structural, and semantic similarity between generated and reference code. Miscellaneous metrics capture operational and cost-related aspects not covered by the other categories.

Tables~\ref{tab:energymetrics} and~\ref{tab:othermetrics} summarize the evaluation metrics reported across the selected studies. Energy consumption is the most commonly reported energy-related metric, while latency/runtime is the dominant performance measure. Fewer studies evaluate functional correctness, memory usage, or code quality. Most of the identified metrics, including energy consumption, power consumption, etc., are conventional software evaluation measures and are not inherently specific to LLM-generated code. In the reviewed studies, their LLM-specific relevance arises from the evaluation context, where these metrics are applied to code generated by different models, prompting strategies, and generation settings. A small number of studies further introduce custom metrics, which are discussed in the next subsection.

\begin{table}[ht!]
\centering

\begin{tabular}{p{2.6cm}|p{5.2cm}}
\hline
\textbf{Energy Metric} & \textbf{Primary Studies} \\
\hline
\hline
Energy consumption &
\cite{peng2024large, rubei2025prompt, cappendijk2025exploration, coignion2024green, cheung2025comparative, rani2025can, vartziotis2024carbon, vartziotis2024learn, islam2025evaluating, podder2025empirical, ashraf2025energy, peng2025sysllmatic, stivala2024investigating, cursaru2024controlled, ashraf2025toward, apsan2025generating} \\

Power consumption &
\cite{coignion2024green, ashraf2025toward} \\

EnergyRatio &
\cite{tuttle2024can} \\

Energy-reduction@k &
\cite{dearing2025leveraging} \\

Carbon footprint &
\cite{cheung2025comparative} \\
\hline

\end{tabular}
\captionsetup{font=small}
\caption{Energy-related evaluation metrics}
\label{tab:energymetrics}
\end{table}

\begin{table}[ht!]
\centering
\renewcommand{\arraystretch}{1.2}
\begin{tabular}{p{1.6cm}|p{2.8cm}|p{3cm}}
\hline
\textbf{Category} & \textbf{Evaluation Metric} & \textbf{Primary Studies} \\
\hline

\multirow{5}{*}{Performance}
& Latency/Runtime &
\cite{peng2024large, rubei2025prompt, cappendijk2025exploration, coignion2024green, cheung2025comparative, rani2025can, vartziotis2024carbon, vartziotis2024learn, islam2025evaluating, podder2025empirical, ashraf2025energy, peng2025sysllmatic, stivala2024investigating, ashraf2025toward} \\ \cline{2-3}

& Runtime Ratio (\%) &
\cite{tuttle2024can} \\ \cline{2-3}

& Throughput (tokens/s) &
\cite{peng2025sysllmatic} \\ \cline{2-3}

& CPU cycles &
\cite{peng2025sysllmatic} \\ \cline{2-3}

& CPU usage (\%) &
\cite{stivala2024investigating} \\
\hline

\multirow{2}{*}{Correctness}
& Functional correctness &
\cite{peng2024large, rani2025can, vartziotis2024carbon, vartziotis2024learn, ashraf2025energy, peng2025sysllmatic} \\ \cline{2-3}

& Pass@k &
\cite{dearing2025leveraging, islam2025evaluating} \\
\hline

\multirow{3}{*}{Memory}
& Memory usage &
\cite{rani2025can, vartziotis2024carbon, vartziotis2024learn, islam2025evaluating, ashraf2025energy, peng2025sysllmatic, stivala2024investigating, ashraf2025toward} \\ \cline{2-3}

& Peak memory &
\cite{cappendijk2025exploration} \\ \cline{2-3}

& FLOPs &
\cite{cappendijk2025exploration, vartziotis2024carbon, vartziotis2024learn} \\
\hline

\multirow{2}{*}{Code Quality}
& Exact Match &
\cite{rubei2025prompt} \\ \cline{2-3}

& AST similarity (\%) &
\cite{cappendijk2025exploration} \\
\hline

\multirow{4}{*}{Miscellaneous}
& Token usage &
\cite{islam2025evaluating} \\ \cline{2-3}

& Monetary cost &
\cite{islam2025evaluating} \\ \cline{2-3}

& Rejected requests &
\cite{coignion2024green} \\ \cline{2-3}

& Completed generations &
\cite{coignion2024green} \\
\hline

\end{tabular}
\captionsetup{font=small}
\caption{Other evaluation metrics reported in the selected studies.}
\label{tab:othermetrics}
\end{table}
\subsubsection{Custom Evaluation Metrics}
In addition to conventional evaluation metrics, a small number of studies introduce custom metrics tailored to specific aspects of energy efficiency and environmental impact. Some of these metrics, particularly RuntimeRatio, EnergyRatio, and Energy-reduction@k, incorporate evaluation characteristics more closely tied to the LLM-generation context, such as comparison with reference solutions or evaluation over multiple generated candidates. Other metrics capture broader energy- and carbon-related characteristics. Table \ref{tab:custom_metrics} summarizes these custom evaluation metrics and the corresponding studies in which they are introduced.

\begin{table}[ht!]
\centering
\renewcommand{\arraystretch}{1.2}
\begin{tabular}{p{0.8cm} p{7cm}}
\hline
\textbf{Study} & \textbf{Custom Evaluation Metrics} \\
\hline
\hline
\cite{tuttle2024can} & RuntimeRatio, EnergyRatio \\
\hline
\cite{dearing2025leveraging} & Energy-reduction@k \\
\hline
\cite{vartziotis2024carbon} & Embodied Energy, Operational Energy, Carbon Intensity, Embodied Carbon Emissions, Operational Carbon Emissions \\
\hline
\cite{vartziotis2024learn} & Green Capacity (GC) \\
\hline
\end{tabular}
\captionsetup{font=small}
\caption{Custom energy-related evaluation metrics.}
\label{tab:custom_metrics}
\end{table}

\subsubsection{Libraries / Tools for Measurement}
To understand how energy is measured in the selected studies, Table~\ref{tab:measurement_tools} presents the tools and libraries used for measuring energy consumption and related system metrics. The table lists each tool as reported by the original authors and maps it to the corresponding primary studies. It can be observed that most studies rely on Python-based measurement libraries or operating system–level tools, such as perf and Intel RAPL. In contrast, only two studies \cite{cursaru2024controlled,apsan2025generating} employ external hardware power monitors (Monsoon Power Monitor), indicating limited use of direct physical energy measurement in the existing literature.

\begin{table}[ht!]
    \centering
    \begin{tabular}{p{4cm} p{3.3cm}}

\hline
\textbf{Tool / Library} & \textbf{Primary Studies} \\
        \hline
\hline
Intel RAPL\cite{intel_sdm_2026} &  \cite{peng2024large, peng2025sysllmatic}\\
CodeCarbon\cite{courty_codecarbon_2026_v321} & \cite{rubei2025prompt, cheung2025comparative, ashraf2025energy, ashraf2025toward}\\
PyJoules\cite{belgaid_pyjoules_2020} & \cite{tuttle2024can} \\
Perf\cite{linux_perf_manpage} & \cite{cappendijk2025exploration, coignion2024green, vartziotis2024learn, islam2025evaluating, stivala2024investigating}\\
PyNVML\cite{nvidia_ml_py_2026} & \cite{dearing2025leveraging} \\
Nvidia-smi\cite{nvidia_smi_docs} &  \cite{coignion2024green}\\
Rocm-smi\cite{rocm_smi_lib_docs_2026} & \cite{dearing2025leveraging} \\
Turbostat\cite{turbostat_manpage} &  \cite{podder2025empirical} \\ 
EnergiBridge\cite{sallou_energibridge_arxiv_2023} &  \cite{rani2025can, apsan2025generating}\\
Windows E3\cite{microsoft_windows_e3_2016} &  \cite{vartziotis2024carbon}\\
Monsoon Power Monitor\cite{monsoon_power_monitor} & \cite{cursaru2024controlled, apsan2025generating}\\
\hline
\end{tabular}
\captionsetup{font=small}
\caption{Tools and libraries used for energy measurement}
\label{tab:measurement_tools}
\end{table}

\subsubsection{Hardware Configuration}
Table~\ref{tab:hardware_eval_category} summarizes the hardware environments used in the selected studies. For clarity and comparability, individual hardware specifications are grouped into high-level categories based on where and how the experiments were executed. On-premise (local machine/server) refers to evaluations conducted on locally owned or institution-managed machines where hardware resources are directly controlled by the authors. Cloud VM \cite{eswaraprasad2017review} / Cloud bare metal \cite{zhang2020high} includes evaluations performed on virtual machines or dedicated servers provided by cloud platforms. Containerized environment refers to experiments executed in container-based setups; for example, \cite{rani2025can} used Docker containers \cite{docker2020docker}. Edge / embedded device \cite{kukunuri2020edgenilm} includes low-power platforms such as Raspberry Pi \cite{upton2016raspberry} devices used in studies \cite{cursaru2024controlled,apsan2025generating}. Finally, not specified refers to studies that do not explicitly report the evaluation hardware, limiting reproducibility and comparability.
\begin{table}[ht!]
\centering
\begin{tabular}{p{4cm}p{3.8cm}}
\hline

\textbf{Evaluation Hardware Category} & \textbf{Primary Studies} \\
\hline
\hline
On-premise (local machine/server) 
& \cite{rubei2025prompt, tuttle2024can, cappendijk2025exploration, coignion2024green, vartziotis2024learn, islam2025evaluating, stivala2024investigating, apsan2025generating} \\
\hline
Cloud 
& \cite{podder2025empirical, ashraf2025energy, peng2025sysllmatic, dearing2025leveraging,ashraf2025toward} \\
\hline
Containerized Environment 
& \cite{rani2025can} \\
\hline
Embedded Device 
& \cite{cursaru2024controlled, apsan2025generating} \\
\hline
Not Specified
& \cite{peng2024large,  cheung2025comparative, vartziotis2024carbon} \\
\hline
\end{tabular}
\captionsetup{font=small}
\caption{Categorization of hardware configurations used for evaluation in the selected studies.}
\label{tab:hardware_eval_category}
\end{table}

\subsubsection{Experimental Configuration}

Table~\ref{tab:inference_settings} summarizes the experimental configuration, including the number of runs, decoding strategy \cite{xia2025tutorial}, temperature \cite{troshin2025control}, maximum token limits, and top-p \cite{nguyen2024turning} and top-k \cite{yerram2024hire} values. The table reports these parameters exactly as described by the original authors to reflect their experimental setups.

Only studies that explicitly report at least one experimental or inference parameter are included in the table. Several studies partially specify their configuration by reporting parameters such as the number of runs or temperature, while leaving other settings unspecified. Overall, the table highlights substantial variation in experimental configurations and reveals that many studies provide limited details on inference settings, which may affect reproducibility and comparability across results.

Accordingly, this comparison is intended to characterize experimental practices and reporting heterogeneity across the literature rather than to support direct efficiency comparisons between studies conducted under different configurations.

\begin{table}[ht!]
    \centering
\begin{tabular}{p{0.7cm}p{0.6cm}p{1.2cm}p{1.5cm}p{1.5cm}p{0.8cm}}
\hline
\textbf{Study} & \textbf{Runs} & \textbf{Decoding Strategy} & \textbf{Temperature} & \textbf{Max Tokens}  & \textbf{Top-p} \\
\hline
\hline
\cite{rubei2025prompt} &  10 & - &  - & - & -  \\ \hline
\cite{tuttle2024can} & 5 & - & - & - & -  \\ \hline
\cite{dearing2025leveraging} & 30 & - & 0.2 & 1024 & -  \\ \hline
\cite{cappendijk2025exploration} & 50 & Greedy & 0.2 & - & -  \\ \hline
\cite{coignion2024green} &  5 & - &  - & - & -  \\ \hline
\cite{cheung2025comparative} &  3 & - &  - & - & -  \\ \hline
\cite{rani2025can} &  30 & - &  0.5 & 4096–8192 & 1  \\ \hline
\cite{vartziotis2024learn} &  10 & - &  - & - & -  \\ \hline
\cite{islam2025evaluating} &  25 & - &  0.7 & - & -  \\ \hline
\cite{ashraf2025energy} &  10 & - &  - & - & -  \\ \hline
\cite{peng2025sysllmatic} &  5 & - & 0.7 & - & -  \\ \hline
\cite{stivala2024investigating} &  10 & - &  - & - & -  \\ \hline
\cite{cursaru2024controlled} &  30 & - &  0.6-1 & - & -  \\ \hline
\cite{ashraf2025toward} &  10 & - &  - & - & -  \\ \hline
\cite{apsan2025generating} &  30 & - &  - & - & -  \\

\hline
\end{tabular}
\captionsetup{font=small}
\caption{Decoding and inference settings reported in the selected studies.}
\label{tab:inference_settings}
\end{table}
\subsubsection{Open-Source Availability}
To assess reproducibility and transparency, we examine the open-source availability of the selected studies. A study is considered open source if it provides publicly accessible code or a replication package. Table \ref{tab:open_source_summary} summarizes the open-source availability of the reviewed studies, indicating whether implementation artifacts are publicly released.

\begin{table}[ht!]
    \centering
    \begin{tabular}{p{3cm} p{4.5cm}}

\hline
\textbf{Category} & \textbf{Primary Studies}\\
        \hline
\hline
Open-source available & \cite{peng2024large, rubei2025prompt, cappendijk2025exploration, coignion2024green,cheung2025comparative, rani2025can, islam2025evaluating, peng2025sysllmatic, stivala2024investigating, cursaru2024controlled, apsan2025generating}\\
Not open-source & \cite{tuttle2024can, dearing2025leveraging, vartziotis2024carbon,   vartziotis2024learn, podder2025empirical, ashraf2025energy,  ashraf2025toward}\\
\hline
\end{tabular}
\captionsetup{font=small}
\caption{Summary of open-source availability}
\label{tab:open_source_summary}
\end{table}

\subsubsection{Benchmark Comparison and Usage}

We categorize the evaluation data based on the type of benchmark or dataset used and the programming language considered. Table~\ref{tab:benchmark_language} summarizes the benchmarks and datasets, along with programming language information when explicitly reported. Programming languages are reported only when explicitly specified by the authors or when the language choice is central to the evaluation; for standardized benchmarks, the programming language is implicit and therefore omitted. Most studies primarily focus on Python, reflecting its widespread use in code generation research. Additionally, not all studies use standardized benchmarks, with several relying on custom data sources, indicating variability in evaluation practices.

\begin{table}[t!]
\centering
\renewcommand{\arraystretch}{1.2}
\begin{tabular}{p{0.6cm}p{4cm}p{3cm}}
\hline
\textbf{Study} & \textbf{Benchmark / Dataset Source} & \textbf{Programming Language} \\
\hline
\cite{peng2024large} & Energy-Language\cite{EnergyLanguagesRepo} & C++ \\
\cite{rubei2025prompt} & CodeXGLUE\cite{Lu2021CodeXGLUE} & -- \\
\cite{tuttle2024can} & LeetCode\cite{LeetCode} & Python \\
\cite{dearing2025leveraging}  & HeCBench suite\cite{jin2023hecbench}, XSBench\cite{forster2015xsbench}, miniMDock\cite{thavappiragasam2022minimdock}& -- \\
\cite{cappendijk2025exploration} & LeetCode & Python \\
\cite{coignion2024green} & CLI Connect 4 Game\cite{CLBG} & Java \\
\cite{cheung2025comparative} & Codeforces\cite{Codeforces} & Python \\
\cite{rani2025can} & GitHub repositories\cite{GitHubAbout} & MATLAB \\
\cite{vartziotis2024carbon} & Not specified & -- \\
\cite{vartziotis2024learn} & LeetCode & Python \\
\cite{islam2025evaluating} & EffiBench\cite{huang2024effibench} & -- \\
\cite{podder2025empirical} & Not Specified & Java \\
\cite{ashraf2025energy} & LeetCode & Python \\
\cite{peng2025sysllmatic} & HumanEval\_CPP\cite{zheng2023codegeex}, SciMark2\cite{SciMark2}, DaCapoBench\cite{Blackburn2006DaCapo} & -- \\
\cite{stivala2024investigating} & HumanEval\cite{chen2021humaneval} & -- \\
\cite{cursaru2024controlled} & LeetCode / HackerRank\cite{HackerRank} & C++, JavaScript, Python \\
\cite{ashraf2025toward} & LeetCode & Python \\
\cite{apsan2025generating} & EvoEval \cite{Xia2024EvoEval} & -- \\
\hline
\end{tabular}
\captionsetup{font=small}
\caption{Benchmarks and programming languages used in the selected studies.}
\label{tab:benchmark_language}
\end{table}

\subsection{Model}

\subsubsection{Model Openness}
To further characterize the experimental settings of the selected studies, we categorized the evaluated models based on their availability and licensing. Studies were classified as closed-source if they exclusively evaluated proprietary or commercial models, open-source if they evaluated only publicly available models, and both open-source and closed-source if they considered a combination of proprietary and open-source models.

\begin{table}[ht!]
    \centering
        \begin{tabular}{p{3cm} p{4cm} p{0.5cm}}
        \hline
        Category & Primary Studies & Total \\
        \hline
        \hline
        Open-Source &  \cite{rubei2025prompt, cappendijk2025exploration, coignion2024green,  cursaru2024controlled, ashraf2025toward} & 5 \\
        Closed-Source&   \cite{peng2024large, cheung2025comparative, vartziotis2024carbon, vartziotis2024learn, podder2025empirical, stivala2024investigating} & 6 \\
        Both Open and Closed-source &  \cite{tuttle2024can, rani2025can, islam2025evaluating, ashraf2025energy, peng2025sysllmatic, apsan2025generating} & 6 \\
        Not Specified &   \cite{dearing2025leveraging} & 1 \\
        \hline
        \end{tabular}
        \captionsetup{font=small}
    \caption{Categorization of studies based on model availability}
    \label{tab:model_availability}
\end{table}

Table~\ref{tab:model_availability} summarizes the categorization of the selected studies based on model openness. It can be seen that the studies are distributed across open-source, closed-source, and mixed model categories, with five studies using open-source models and six studies each in the closed-source and mixed categories. In addition, although one study \cite{dearing2025leveraging} discusses the use of proprietary models, it does not clearly specify which models are evaluated and is therefore categorized as not specified.

\begin{table*}[ht!]
\centering
\scriptsize
\begin{tabular}{p{1cm} p{12.5cm} p{3cm}}
\hline
\textbf{Study} & \textbf{Models Evaluated} & \textbf{Model Size Category} \\
\hline
\hline
\cite{peng2024large} & GPT-4o & Large \\
\cite{rubei2025prompt} & Llama 3 8B-Instruct & Medium \\
\cite{tuttle2024can} & ChatGPT-3.5, ChatGPT-4o, CodeGemma-7B, WizardCoder-33B, Llama 3-8B, Phind-CodeLlama-34B-v2, Nous-Hermes 2-10.7B, Mistral-7B & Mixed (Small + Large) \\
\cite{dearing2025leveraging} & Not Specified & Not specified \\
\cite{cappendijk2025exploration} & Code Llama-70B, Code Llama-70B-Instruct, Code Llama-70B-Python, DeepSeek-Coder-33B-Base, DeepSeek-Coder-33B-Instruct & Large \\
\cite{coignion2024green} & GitHub Copilot (StarCoder family back-end models: StarCoder 15.5 B, StarCoder2-7 B, StarCoder2-15 B) & Mixed (Small + Medium) \\
\cite{cheung2025comparative} & GPT-4 & Large \\
\cite{rani2025can} & GPT-3, GPT-4, Meta Llama-3 70B Instruct, Mixtral 8$\times$22B Instruct & Large \\
\cite{vartziotis2024carbon} & GitHub Copilot & Not specified \\
\cite{vartziotis2024learn} & GitHub Copilot, OpenAI ChatGPT-3, Amazon CodeWhisperer & Large \\
\cite{islam2025evaluating} & Amazon Nova-Lite, Amazon Nova-Micro, Amazon Nova-Pro, Anthropic Claude 3.5 Haiku, Anthropic Claude 3.5 Sonnet, DeepSeek v3 (37B), Google Gemini 1.5 Flash, Google Gemini 1.5 Pro, Google Gemini 2.0 Flash, Google Gemini 2.0 Flash-Lite, Meta Llama 3.1 (8B), Meta Llama 3.1 (70B), Meta Llama 3.3 (70B), Mistral Codestral-Mamba-2407 (7B), Mistral-Large-2407 (123B), Pixtral-Large-2411 (124B), OpenAI GPT-3.5 Turbo, OpenAI GPT-4 Turbo, OpenAI GPT-4o, xAI Grok & Mixed (Small + Medium + Large) \\
\cite{podder2025empirical} & GitHub Copilot & Not specified \\
\cite{ashraf2025energy} & StableCode-3B, StarCoderBase-3B, Qwen2.5-Coder-3B-Instruct, GPT-4.0, DeepSeek-Reasoner & Mixed (Small + Large) \\
\cite{peng2025sysllmatic} & GPT-4o, GPT-4.1, qwen3-coder:480b, gemma3:27b, deepseek-r1:70b, llama4:latest & Mixed (Medium + Large) \\
\cite{stivala2024investigating} & GitHub Copilot & Not specified \\
\cite{cursaru2024controlled} & CodeLlama-7B-Instruct & Small \\
\cite{ashraf2025toward} & StableCode-Instruct-3B, Qwen2.5-Coder-3B-Instruct, CodeLlama-7B-Instruct, Phi-3-Mini-4K-Instruct & Small \\
\cite{apsan2025generating} & ChatGPT (175B), GPT-4 (1.76T), DeepSeek Coder Instruct (33B), Speechless CodeLlama v2.0 (34B), CodeMillennials (34B), WizardCoder v1.1 (33B) & Large \\
\hline
\end{tabular}
\captionsetup{font=small}
\caption{Overview of the evaluated models and their parameter-scale categories.}
\label{tab:models-evaluated}
\end{table*}

\subsubsection{Model Domain Specialization}
To characterize the domain focus of the evaluated models, we classify them as general-purpose, code-specialized, or mixed (general and code-specialized) language models. General-purpose LLMs \cite{wang2024theoremllama} are designed for a broad range of natural language tasks without domain-specific optimization. Code-specialized LLMs \cite{anand2024critical} are trained or fine-tuned for programming-related tasks, including code generation, completion, and analysis. Studies that evaluate both general-purpose and code-specialized models are classified as mixed. Studies that do not report the domain specialization of the evaluated models are labeled as not specified.

\begin{table}[ht!]
    \centering
    \begin{tabular}{p{3cm} p{4cm} p{0.5cm}}
        \hline
        \textbf{Category} & \textbf{Primary Studies} & \textbf{Total} \\
        \hline
        \hline
        General-Purpose LLMs & \cite{peng2024large, rubei2025prompt, cheung2025comparative, rani2025can} &  4\\
        Code-Specialized LLMs & \cite{cappendijk2025exploration, coignion2024green,  vartziotis2024carbon, podder2025empirical, stivala2024investigating, cursaru2024controlled, ashraf2025toward} & 7 \\
        Mixed & \cite{tuttle2024can, vartziotis2024learn, islam2025evaluating, ashraf2025energy, peng2025sysllmatic,  apsan2025generating} &  6\\
        Not Specified &  \cite{dearing2025leveraging} & 1 \\
        \hline
    \end{tabular}
    \captionsetup{font=small}
    \caption{Categorization of studies based on model domain specialization}
    \label{tab:domain}
\end{table}

The distribution of studies across these domains is presented in Table \ref{tab:domain}. It can be seen that only four studies focus exclusively on general-purpose language models, while the remaining studies evaluate either code-specialized models or include at least one code-specialized model in a mixed evaluation setting.

\subsubsection{Evaluated Models and Model Size Categories}
We first present the list of models considered in each study in Table~\ref{tab:models-evaluated}. The models are listed as reported by the original authors to accurately reflect the experimental settings used in each work. To enable consistent comparison, we categorize models by parameter count into three groups: small ($\leq 7B$), medium (8–30B), and large ($ > 30B$). Some studies evaluate models spanning multiple size categories and are therefore labelled as mixed-scale to accurately reflect their evaluation scope. Studies that do not report model sizes or rely on proprietary systems with undisclosed back-end configurations are labelled as not specified. 

It should be noted that these model-size categories are used only to provide a descriptive overview of the models evaluated in the literature. They do not imply that parameter size alone determines energy efficiency or performance. Models within the same size category may differ in architecture, training data, tokenizer, context length, quantization, decoding settings, and deployment environment. Therefore, these categories are used to summarize model-usage trends rather than to directly compare energy efficiency across model sizes.

\subsection{Model Adaptation}
\subsubsection{Fine-Tuning}
We categorize the studies based on the use of fine-tuning, i.e., whether model parameters are adapted to generate more sustainable code. Notably, none of the reviewed studies employ fine-tuning specifically for improving energy-efficiency. This highlights an open research direction in leveraging fine-tuning to guide LLMs toward more energy-efficient and resource-aware code generation.

\subsubsection{Prompt Engineering}
Table~\ref{tab:prompt_engineering} summarizes the use of prompt engineering techniques in the selected studies. A study is marked as “Yes” if it explicitly applies methods such as zero-shot, few-shot, chain-of-thought, or prompt variations, and as “No” if default prompts are used. Several studies employ prompt-based strategies, while others evaluate models without specific prompt engineering.
\begin{table}[ht!]
\centering
\footnotesize
\begin{tabular}{p{2.5cm} p{5cm}}
\hline
\textbf{Prompt Engineering} & \textbf{Primary Studies} \\
\hline
\hline
Yes & 
\cite{peng2024large, 
rubei2025prompt, 
dearing2025leveraging, 
cappendijk2025exploration, 
podder2025empirical, 
peng2025sysllmatic, 
cursaru2024controlled, 
ashraf2025toward, 
apsan2025generating,
tuttle2024can,rani2025can, stivala2024investigating}
\\
\hline
No & 
  
\cite{cheung2025comparative, 
vartziotis2024carbon, 
vartziotis2024learn, 
islam2025evaluating, 
ashraf2025energy, 
coignion2024green} \\
\hline
\end{tabular}
\captionsetup{font=small}
\caption{Use of prompt engineering techniques}
\label{tab:prompt_engineering}
\end{table}
\section{Discussions and Synthesis}
\label{subsec:results_rqs}

\subsection{RQ1 Analysis}
To systematically synthesize the findings across the selected studies, we conducted an inductive thematic analysis of the reported energy-efficiency outcomes of LLM-generated code. The identified themes are discussed below.

\textbf{T1 - LLM-generated code does not consistently outperform human written code:}
Among the 18 reviewed studies, eight studies (\cite{vartziotis2024learn,tuttle2024can,cheung2025comparative,rani2025can,islam2025evaluating,cursaru2024controlled,apsan2025generating, cappendijk2025exploration}) report that LLM-generated code does not outperform human baselines, while \cite{ashraf2025energy} report mixed outcomes across tasks and experimental settings. The remaining nine studies (\cite{peng2024large,rubei2025prompt,dearing2025leveraging,coignion2024green,podder2025empirical,peng2025sysllmatic,stivala2024investigating,ashraf2025toward,vartziotis2024carbon}) evaluate energy-efficiency without directly comparing LLM-generated and human-written solutions. Tuttle et al. \cite{tuttle2024can} find that LLM-generated solutions outperform the human baseline in only 11 of 192 responses (5.7\%). Similarly, Islam et al. \cite{islam2025evaluating} report that the best performing LLM, DeepSeek-v3, consumes 6.37 J compared with 5.46 J for the human baseline on the larger benchmark. In contrast, Ashraf et al. \cite{ashraf2025energy} report mixed results across model sizes. At least one SLM matches or outperforms the LLM/human baseline on 79 of 150 problems (52.6\%), while larger models generally achieve higher correctness and faster execution. Vartziotis et al. \cite{vartziotis2024carbon} identify a different trade-off, in which Copilot reduces the estimated development time from approximately 78 hours to just over 5 hours. However, LLM inference consumes 9.203 kWh compared with 1.131 kWh for executing the generated code. These findings show that LLM-generated code can provide energy-efficiency benefits in specific settings, but a consistent advantage over human-written solutions has not been established. The quality assessment generally supports the reliability of the reported comparisons, as QA6 indicates that 17 of the 18 studies use appropriate baselines and evaluation measures. However, the type of baseline varies across studies and does not always involve human-written code. Therefore, the conclusions regarding the relative energy efficiency of LLM-generated and human-written code are drawn from the nine studies that directly compare these two types of solutions.

\textbf{T2 - Energy-efficiency outcomes are model dependent, but model size alone is not a reliable predictor:} 
Several studies (\cite{tuttle2024can,cappendijk2025exploration,coignion2024green,rani2025can,islam2025evaluating,ashraf2025toward,ashraf2025energy,peng2024large,apsan2025generating}) report differences in energy-efficiency outcomes across models. However, only a smaller subset directly examines the effect of model size. The remaining studies either evaluate a single model or do not directly compare small and large models. Islam et al. \cite{islam2025evaluating} find that DeepSeek-v3 (37B) achieves better energy performance than several larger models, including 70B Llama variants. Ashraf et al. \cite{ashraf2025energy} further show that correctly generated code from SLMs can sometimes match or outperform larger models and human-written solutions in terms of energy efficiency. In contrast, Peng et al. \cite{peng2024large} find that larger models provide stronger optimization capabilities for complex software systems. Ashraf et al. \cite{ashraf2025toward} also report differences among models of similar size, with Qwen2.5-Coder-3B outperforming other SLMs under several prompting strategies. These observations suggest that the energy efficiency of generated code is influenced not only by model size, but also by factors such as model architecture, specialization, and task compatibility.

\textbf{T3 - Energy efficiency is strongly task- and algorithm-dependent:}
Among the reviewed studies, fourteen studies (\cite{vartziotis2024learn, peng2024large, tuttle2024can, dearing2025leveraging, cappendijk2025exploration, cheung2025comparative, rani2025can, islam2025evaluating, podder2025empirical, ashraf2025energy, peng2025sysllmatic, stivala2024investigating, cursaru2024controlled, ashraf2025toward}) provide evidence that energy efficiency outcomes vary across programming tasks, workloads, and algorithmic choices. Islam et al. \cite{islam2025evaluating} report relatively competitive energy performance for Breadth-First-Search (BFS), Depth-First-Search(DFS), Binary Search, and Two-Pointer problems, whereas Dynamic Programming, Backtracking, Bit Manipulation, and Greedy problems exhibit substantially higher energy consumption. Cappendijk et al. \cite{cappendijk2025exploration} further show that energy savings depend on the implementation generated for a given problem, particularly the choice of algorithms and library functions. Similarly, Podder et al. \cite{podder2025empirical} demonstrate that appropriate algorithmic optimizations can reduce energy consumption, whereas unsuitable optimization choices can increase it. These observations suggest that the energy efficiency of generated code is strongly influenced by the characteristics of the programming task and the algorithmic decisions reflected in the resulting implementation.

\textbf{T4 - Energy efficiency must be considered together with functional correctness:}
The relationship between energy efficiency and correctness is examined in fifteen of the reviewed studies. Eight studies \cite{peng2024large,rubei2025prompt,dearing2025leveraging,islam2025evaluating,podder2025empirical,peng2025sysllmatic,stivala2024investigating,apsan2025generating} report cases where energy efficient outputs are also functionally correct. In contrast, five studies \cite{tuttle2024can,cheung2025comparative,rani2025can,vartziotis2024carbon,ashraf2025toward} show that lower energy solutions are not always correct, while two studies \cite{vartziotis2024learn,ashraf2025energy} report mixed outcomes. This pattern shows that lower energy consumption alone is not sufficient to assess sustainable code generation. Islam et al. \cite{islam2025evaluating} demonstrate that even functionally correct LLM-generated code can consume more energy than human-written solutions. These results suggest that energy-efficiency should be evaluated as a multi-objective problem, where energy savings are considered together with functional correctness.

QA3 provides important context for interpreting this theme. Thirteen studies validate the functional correctness of the generated code before comparing its energy efficiency, and two studies provide only partial validation. Therefore, the findings of this theme are supported more strongly by studies that explicitly validate the generated code, whereas studies with only partial correctness validation provide weaker evidence for the relationship between energy efficiency and correctness.

\textbf{T5 - Energy efficiency varies across programming languages, but the evidence remains language-specific:}
Programming language also affects the energy efficiency of LLM-generated code, but the results differ across tasks and execution settings. Cursaru et al. \cite{cursaru2024controlled} directly compare C++, JavaScript, and Python using CodeLlama-7B. Their results show that LLM-generated JavaScript consumes less energy than the corresponding human-written implementation for all three evaluated tasks. In contrast, generated C++ and Python code is substantially less efficient than the human baseline for the CPU-bound tasks, with Python showing particularly high energy consumption. The same study also reports that energy-aware prompting improves C++ Two Sum in one case, but increases energy consumption for all three Python tasks.

Java-based studies also report strong but task-specific improvements. Podder et al. \cite{podder2025empirical} evaluate Java code and report energy reductions of up to 94.90\% for sorting and 99.84\% for repeated searching through algorithmic optimization. Peng et al. \cite{peng2025sysllmatic} report a 1.66$\times$ energy improvement on Java SciMark2 kernels and 2.23$\times$ on BioJava, but the improvements are not consistent across all applications. In MATLAB, Rani et al. \cite{rani2025can} find no statistically significant reduction in execution energy across the evaluated LLMs, indicating limited and inconsistent benefits for MATLAB code optimization. The evidence indicates that programming language influences energy efficiency, but its impact depends on the task, algorithm, and implementation. No programming language is consistently the most energy-efficient across the reviewed studies.

\begin{tcolorbox}[
colback=gray!10,
colframe=gray!60,
title=\textbf{Summary of RQ1},
boxrule=0.6pt,
arc=2pt
]
The reviewed studies show that LLM-generated code does not consistently outperform human-written code in energy efficiency. Outcomes vary substantially across models, tasks, algorithms, programming languages, and execution settings, while model size alone is not a reliable predictor of energy performance. Energy efficiency must be evaluated alongside functional correctness, as lower energy use does not guarantee correct code. Overall, no model, prompting approach, or programming language is consistently the most energy-efficient across all settings.
\end{tcolorbox}

\subsection{RQ2 Analysis}
This research question investigates how energy-efficiency is evaluated in the literature by identifying the most commonly used metrics such as energy consumption, execution time, memory usage, carbon footprint, etc. The reviewed studies predominantly rely on conventional software and systems metrics to evaluate the energy efficiency of LLM-generated code. Across the 18 selected studies in Table \ref{tab:energymetrics}, energy-efficiency evaluation is dominated by conventional software and systems metrics. Energy consumption is the most widely used metric, reported in 16 studies; among these, two studies also report power consumption and one reports carbon footprint. These measures are not specific to LLM-generated code, but are applied to generated programs under different models, prompts, and execution settings. Several studies also report complementary measures such as execution time, memory usage, throughput, and functional correctness to characterize the efficiency and validity of generated code, as shown in Table. \ref{tab:othermetrics}. These metrics are not inherently specific to LLM-generated code; their relevance arises from applying them to code produced under different models, prompts, generation settings, and execution environments.

Finally, 4 studies introduce custom metrics (\cite{vartziotis2024learn}, \cite{tuttle2024can}, \cite{dearing2025leveraging}, \cite{vartziotis2024carbon}) that are more closely tailored to the LLM-generation context as shown in Table \ref{tab:custom_metrics}. RuntimeRatio and EnergyRatio compare the runtime and energy consumption of model-generated code against canonical human-written solutions, providing a relative efficiency measure. Energy-reduction@k estimates the expected energy savings when generating multiple candidate solutions and selecting the most energy-efficient valid one, while accounting for both performance improvement and correctness probability. Similarly, Green Capacity measures the overall energy improvement of optimized code across multiple performance dimensions, filtering out invalid solutions. Other metrics, including Embodied Energy, Operational Energy, Embodied Carbon Emissions, Operational Carbon Emissions, and Carbon Intensity, extend the evaluation beyond code execution to broader environmental impacts.

\begin{tcolorbox}[
colback=gray!10,
colframe=gray!60,
title=\textbf{Summary of RQ2},
boxrule=0.6pt,
arc=2pt
]
In summary, the reviewed studies primarily use conventional software and systems metrics to evaluate the energy efficiency of LLM-generated code. Only four studies introduce custom metrics tailored to the LLM-generation context, indicating limited development of LLM-specific energy-efficiency measures.
\end{tcolorbox}

\subsection{RQ3 Analysis}
This research question investigates the benchmarks, datasets, programming languages, and energy 
measurement tools used to evaluate the energy efficiency of LLM-generated code, and examines the extent to which these evaluation resources are specifically designed for energy efficiency assessment.
Across the 18 selected studies, benchmark usage is heterogeneous, with studies relying on programming challenge platforms, established code generation benchmarks, domain specific benchmark suites, custom datasets, and real-world code repositories, as summarized in Table~\ref{tab:benchmark_language}.  

Programming challenge platforms represent the most common source of evaluation problems, with 7 of
the 18 studies using LeetCode, Codeforces, or HackerRank
(\cite{tuttle2024can, cappendijk2025exploration, cheung2025comparative, vartziotis2024learn, ashraf2025energy, cursaru2024controlled, ashraf2025toward}). These studies typically select or construct subsets of algorithmic programming problems and use them to evaluate the correctness and energy efficiency of LLM-generated solutions. A further 4 studies employ established code-generation benchmarks, including
CodeXGLUE, HumanEval, EvoEval, and EffiBench
(\cite{rubei2025prompt,islam2025evaluating, stivala2024investigating, apsan2025generating}). Two
studies use domain specific benchmark suites, including HeCBench, XSBench, miniMDock, SciMark2, and
DaCapoBench (\cite{dearing2025leveraging, peng2025sysllmatic}), while one study evaluates real-world
software using GitHub repositories \cite{rani2025can}. In addition, two studies introduce custom
evaluation resources, namely the Energy-Language dataset and a CLI-based Connect-4 application
(\cite{peng2024large, coignion2024green}), whereas two studies do not clearly specify the dataset used
(\cite{vartziotis2024carbon, podder2025empirical}). Overall, the distribution indicates that most studies
reuse or adapt existing programming and code generation resources by incorporating energy measurements, rather than relying on a commonly adopted benchmark specifically designed for evaluating the energy efficiency of LLM-generated code.

With respect to programming languages, the reviewed experiments are concentrated in a relatively small number of languages. Among the studies for which the programming language is explicitly
reported in Table~\ref{tab:benchmark_language} , Python appears in 7 studies, making it the most frequently evaluated language
(\cite{tuttle2024can, cappendijk2025exploration, cheung2025comparative, vartziotis2024learn, ashraf2025energy, cursaru2024controlled, ashraf2025toward}). C++ is evaluated in 2 studies (\cite{peng2024large,cursaru2024controlled}) and Java in 2 studies
(\cite{coignion2024green, podder2025empirical}), while JavaScript
and MATLAB each appear in only 1 study (\cite{cursaru2024controlled,rani2025can}). Some studies using standardized benchmarks do not explicitly report the programming language and are therefore excluded
from these counts. Other programming languages, such as Go and Solidity, are not represented in the selected studies. This limited language diversity may restrict the generalizability of the findings because the energy characteristics of generated programs can vary across programming languages, runtime
systems, compilers, and execution environments.

Energy-measurement practices are also dominated by software- and system-level tools, as summarized in Table~\ref{tab:measurement_tools}. Seventeen of the 18 studies employ at least one software- or system-level
measurement approach, including Intel RAPL, CodeCarbon, PyJoules, Perf, PyNVML, Nvidia-smi, Rocm-smi, Turbostat, EnergiBridge, and Windows E3.
Among the individual tools, Perf
is the most frequently used, appearing in 5 studies
(\cite{cappendijk2025exploration, coignion2024green, islam2025evaluating, vartziotis2024learn,stivala2024investigating}), followed by CodeCarbon in 4 studies
(\cite{rubei2025prompt, cheung2025comparative, ashraf2025toward, ashraf2025energy}). Intel RAPL (\cite{peng2024large,peng2025sysllmatic}) and EnergiBridge
(\cite{rani2025can,apsan2025generating}) are each used in 2 studies, while PyJoules
\cite{tuttle2024can}, PyNVML and Rocm-smi \cite{dearing2025leveraging}, Nvidia-smi
\cite{coignion2024green}, Turbostat \cite{podder2025empirical}, and Windows E3
\cite{vartziotis2024carbon} each appear in one study. These approaches obtain energy and power related information through processor counters, operating-system interfaces, hardware telemetry, or software-based estimation. However, only 2 of the 18 studies use an external Monsoon
Power Monitor (\cite{cursaru2024controlled,apsan2025generating}), in Raspberry Pi-based experimental
settings. This limited use of external physical measurement indicates that the current literature predominantly depends on software accessible energy measurements rather than direct power 
monitoring instrumentation. Overall, the variation in benchmarks, programming languages, and energy measurement tools indicates limited standardization across existing evaluations, which can reduce the
comparability and reproducibility of reported energy-efficiency results.

\begin{tcolorbox}[
colback=gray!10,
colframe=gray!60,
title=\textbf{Summary of RQ3},
boxrule=0.6pt,
arc=2pt
]
The reviewed studies use a diverse set of benchmarks, datasets, and measurement tools, with limited standardization across evaluations. Evaluations are dominated by Python, C++, and Java, with limited coverage of other programming languages. Energy measurement primarily relies on software-based tools, while direct hardware-level measurement using external power monitors is rare.

\end{tcolorbox}  

\subsection{RQ4 Analysis}
To further investigate beyond the comparison of individual prompting techniques based on frequency, we conducted a qualitative thematic synthesis of findings on prompting and model adaptation. Our coding identified five themes: (T1) Explicit energy-efficient instructions do not always generate greener code, (T2) Structured prompting provides conditional benefits, (T3) Iterative and feedback-guided optimization shows stronger potential than static prompting, (T4) Prompt effectiveness is strongly context-dependent and, (T5) Energy-aware fine-tuning remains underexplored.

\textbf{T1 - Explicit energy-aware instructions do not always generate greener code:} Studies using direct energy-aware prompts report mixed results. Tuttle et al., \cite{ tuttle2024can}, found that energy-aware prompting did not improve energy efficiency across all eight LLMs. In some cases, re-prompting achieved substantial savings, including a reduction from 18.334 J to 0.994 J. Rani et al., \cite{ rani2025can}, found no statistically significant improvement in energy consumption across the evaluated models. Stivala et al., \cite{ stivala2024investigating}, reported an average energy reduction of approximately 18\%, but the difference was not statistically significant. Cursaru et al., \cite{cursaru2024controlled}, observed that explicitly requesting energy efficient code did not consistently reduce energy consumption, with the results varying across programming tasks and languages. Overall, the evidence indicates that direct energy-aware prompting can reduce energy consumption in specific cases, but its effectiveness is inconsistent across models, tasks, and programming languages.

\textbf{T2 - Structured prompting provides conditional benefits: } 
The reviewed literature further investigates whether additional structure such as few-shot prompting, role-based prompting, CoT, re-prompting, and structured instructions  can improve the energy efficiency of generated code. Zero-shot prompting does not consistently reduce energy consumption and, in some cases, performs same as or worse than the baseline approaches \cite{tuttle2024can,vartziotis2024learn}. In comparison, CoT prompting shows more favorable results for selected models. Ashraf et al. \cite{ashraf2025toward} find that CoT reduces energy consumption for Qwen2.5-Coder-3B and StableCode-3B, whereas CodeLlama-7B and Phi-3-Mini do not outperform the baseline under the evaluated prompting strategies.
Cappendijk et al. \cite{cappendijk2025exploration} report substantial execution energy reductions for selected model–prompt–problem combinations, including 59.7\% for a for-loop-oriented prompt and 59.4\% for a library-function prompt with CodeLlama-70B-Python. However, the same study also reports a 465.5\% increase in energy consumption, showing that structured optimization instructions can also lead to less energy efficient code.  Rubei et al. \cite{rubei2025prompt} showed that structured prompts using explanatory tags can reduce LLM inference energy compared with default prompt configurations, but their analysis does not consider the execution energy of the generated code. The evidence suggests that structured prompting can improve energy efficiency for particular model--prompt--problem combinations, but greater prompt complexity does not consistently translate into lower energy consumption.

\textbf{T3 - Iterative and feedback-guided optimization shows stronger potential than static prompting:}
Iterative approaches use execution feedback, energy measurements, profiling data, and automated evaluation to improve the energy efficiency of generated code. Peng et al.,\cite{peng2024large} combine energy-aware prompting with evaluator feedback and report energy reductions in 83\% of the six successfully evaluated programs, including a reduction from 31 J to 15 J for the spectral-norm benchmark. Podder et al., \cite{podder2025empirical} use iterative green code prompts that are validated by developers and report a 35.85\% reduction in execution time across five AWS Lambda hotspots, which is used to estimate a corresponding reduction in energy consumption and carbon emissions. Dearing et al., \cite{dearing2025leveraging} propose LASSI-EE, which integrates runtime power profiling, hardware-aware context, iterative refinement, and energy-based candidate selection. LASSI-EE achieves an expected energy reduction of 29\% with a single candidate and 48\% with three candidates, as well as a 2.8$\times$ greater expected energy reduction than vanilla prompting in the single-candidate setting. Collectively, these findings suggest that incorporating empirical execution feedback, profiling information, and iterative candidate refinement can be more effective for energy optimization than relying solely on static energy-aware instructions.

\textbf{T4 -  Prompt effectiveness is strongly context-dependent:}
The energy impact of a prompting strategy varies with the model, programming task, generated implementation, language, and execution environment. Cappendijk et al.  
\cite{cappendijk2025exploration} show that for-loop optimization prompts can reduce energy consumption when they lead the model to select a more efficient library based sorting method. In contrast, energy consumption can increase when the generated code uses a less efficient algorithm. Apsan et al. \cite{apsan2025generating} report that the most energy efficient prompting strategy varies across hardware platforms. Few shot prompting performs best on the server, keyword prompting on the PC, and guideline based prompting on Raspberry Pi. Human-written solutions are 16\% more energy-efficient on the server and 3\% on Raspberry Pi, whereas LLM-generated solutions are 25\% more energy-efficient on the PC. These results highlight the influence of the execution environment on the energy performance of generated code. Cursaru et al. \cite{cursaru2024controlled} also report that the effect of prompting varies across programming languages and benchmark problems. Overall, prompt effectiveness emerges from the interaction among the model, programming task, generated implementation, language, and execution environment rather than from the prompting strategy alone.

The quality assessment provides relatively high confidence in the reporting of the execution environment and prompting procedures. QA4 shows that 15 of the 18 studies sufficiently describe the hardware and execution environment, while QA9 indicates that prompting and generation settings are sufficiently reported in 16 studies. In contrast, QA5 provides only moderate confidence in the reporting of broader experimental and generation settings, since only six studies provide complete information. Therefore, the finding that prompt effectiveness is context-dependent is well supported by the reported execution environments and prompting procedures, although comparisons across studies should be made with some caution because important generation parameters are not consistently reported.

\textbf{T5 - Energy-aware fine-tuning remains underexplored:}
Although RQ4 considers both prompting and model-adaptation methods, the reviewed literature is dominated by inference-time prompting strategies. Fine-tuning approaches designed to generate more energy-efficient code remain largely unexplored, limiting direct comparison between prompt engineering and model adaptation. This indicates a clear need for future work on energy-aware fine-tuning and its effectiveness across models, tasks, and programming languages.

\begin{tcolorbox}[
    colback=gray!10,
    colframe=gray!60,
    title=\textbf{Summary of RQ4},
    boxrule=0.6pt,
    arc=2pt
]
The reviewed studies show that prompting can improve the energy efficiency of LLM-generated code, but its effectiveness is inconsistent and context-dependent. Direct energy-aware prompts do not reliably produce greener code, while structured and iterative feedback-guided approaches show greater potential in specific settings. No single prompting strategy consistently performs best across models, tasks, languages, and hardware environments. Moreover, energy-aware fine-tuning and other model-adaptation methods remain largely underexplored.
\end{tcolorbox}

\subsection{Implications for Industry Practitioners}
While RQ1--RQ4 characterize the current state of research on energy-efficient LLM-generated code, the findings also provide direct implications for practitioners who use LLMs in software development. First, practitioners should not assume that LLM-generated code is energy-efficient by default. The findings in RQ1 show that generated solutions outperform human baselines in only a small number of cases. Therefore, energy efficiency should be verified through profiling rather than assumed as a benefit of using LLMs. Model selection should also depend on the task rather than model size alone. Smaller models have matched or outperformed larger models on some problems, while larger models are more effective mainly for complex optimization tasks. Therefore, organizations may benefit from evaluating smaller and less expensive models on representative internal workloads before selecting the largest available model. Since energy outcomes are also influenced by the algorithmic and library choices in the generated code, prompts can be designed to encourage known energy-efficient patterns for a specific language.

Energy efficiency should also not be considered separately from correctness. The most energy-efficient candidate is not always functionally correct, so energy-aware generation pipelines should first verify correctness before comparing energy performance. Similarly, because no programming language is consistently the most energy-efficient across studies, practitioners working in energy-sensitive environments should evaluate generated code using their own programming language and runtime rather than generalizing from results obtained in other environments.

For prompting, simple energy-aware instructions do not consistently improve energy efficiency. In contrast, iterative and feedback-guided approaches that use execution profiling and automated candidate selection show stronger results, including energy reductions of 29--48\% and improvements in more than 80\% of evaluated cases. Organizations with existing CI/CD infrastructure can therefore integrate energy measurement into a generate--measure--refine process rather than relying only on prompt engineering.

Finally, the effectiveness of prompting strategies can vary across deployment environments. Results obtained on one hardware platform should therefore not be assumed to apply directly to another, and re-validation is recommended when the target environment changes. Overall, these considerations translate the findings of RQ1--RQ4 into practical guidance for industry teams making evidence-based decisions when adopting LLMs for sustainable software development.
\section{Future Research Opportunities}

\subsection{Expanding Energy-Efficiency Beyond Code Generation}

One notable observation from our review is that energy-efficiency evaluation remains focused mainly on initial code generation. As shown in Table~\ref{tab:code-related}, relatively few studies examine code optimization, completion, or refactoring, limiting our understanding of how LLMs can support energy efficiency beyond code generation.

Future research should therefore examine how LLMs can support energy-aware code optimization and refactoring. An important question is whether LLMs can identify energy-intensive code regions, inefficient patterns, performance antipatterns, or code smells that lead to unnecessary computation, memory usage, data movement, or I/O. Existing knowledge from green software engineering can guide this direction by linking known energy-related patterns with LLM-based code analysis and transformation. Another promising direction is whether LLMs can recommend or automatically apply energy-efficient transformations while preserving code correctness, such as selecting better algorithms and data structures, removing redundant computations, reducing unnecessary memory allocations, minimizing expensive I/O, or replacing inefficient API usage. For example, an LLM could identify a computational or memory bottleneck and generate an optimized implementation using energy, runtime, or resource feedback.

Energy-aware refactoring also deserves greater attention. Refactoring techniques are commonly used to improve code structure and maintainability, but future studies could investigate whether LLM-assisted refactoring can additionally reduce energy consumption without introducing functional regressions. Similarly, code completion systems could be designed to recommend more efficient implementation choices during development, helping developers avoid inefficient patterns before they become part of the final software. More generally, future work should examine whether LLMs can \emph{detect, explain, and repair} energy inefficiencies in existing software. Combining LLMs with profiling, static analysis, energy measurements, and known energy-related code patterns could support practical tools for continuous energy optimization and maintenance.

\subsection{Underexplored Application Domains}

An important gap identified in this review is the limited exploration of energy-aware code generation across diverse application domains. The majority of the selected studies focus on general-purpose software development, with only two studies addressing scientific and engineering applications. This distribution indicates that energy-aware research in LLM-based code generation is primarily centered on standard programming tasks and general development scenarios. 

Future research should explore energy-aware code generation across a broader range of application domains. In particular, mobile and client-side applications, low-level and systems software, web and cloud infrastructure, and IoT and embedded systems are not adequately represented in the selected studies. These domains are especially important from an energy-efficiency perspective, as energy efficiency directly affects battery life, system performance, scalability, and overall environmental impact. For example, embedded and IoT systems operate under strict energy constraints \cite{s151024818}, low-level software can significantly influence hardware efficiency \cite{8130281}, and cloud-based services contribute substantially to large-scale energy consumption \cite{doi:10.1126/science.aba3758}. The limited attention to these areas highlights the need for future studies to evaluate LLM-generated code in more realistic, energy-sensitive, and deployment-oriented environments.

\subsection{Expanding Energy-Efficiency Evaluation Across Programming Languages}
Prior work has shown that programming languages and execution environments can exhibit different energy and performance characteristics \cite{10.1145/3136014.3136031, PEREIRA2021102609}. However, it remains unclear whether these differences are consistently reflected in LLM-generated code. Our review shows that this diversity is not yet well represented in studies evaluating the energy efficiency of LLM-generated code. Most selected studies focus on Python, Java, and C++, with only limited attention to other languages and programming environments.

This gap is important because the energy efficiency of generated code can depend on language-specific abstractions, memory management, runtime behavior, compiler support, and the target execution environment \cite{georgiou2019software}. Although these factors are already well studied in conventional software-energy research, their influence on LLM-generated code remains underexplored. Future studies should therefore examine whether LLMs can consistently generate energy-efficient solutions across a broader range of programming languages and whether the observed efficiency patterns generalize across different execution environments.

In particular, low-level and system-oriented languages such as C and Rust provide opportunities to investigate whether LLMs can generate efficient implementations that make effective use of memory, concurrency, and hardware resources. Similarly, Go is relevant for cloud-native systems, JavaScript for client-side applications, Swift and Kotlin for mobile development, and Solidity and Rust for blockchain applications, where inefficient generated code can lead to higher computational or transaction costs. Domain-specific languages used in scientific computing, data engineering, and other specialized environments also remain largely unexplored. Expanding evaluation across diverse programming ecosystems would clarify how language and execution environment affect the energy efficiency of LLM-generated code and whether current findings generalize beyond commonly studied languages.

\subsection{Toward energy-Aware LLM Optimization}
Another important direction for future research is the development of sustainability-aware fine-tuning and model adaptation techniques. Our review shows that no study used model adaptation techniques, such as fine-tuning, with an explicit focus on improving energy efficiency. This suggests that most existing work measures energy at the output level without modifying the underlying model to generate more energy-efficient code.

Future research should explore energy-aware fine-tuning strategies specifically tailored for sustainable code generation and optimization tasks. Instead of only measuring the energy consumption of generated code after inference, models should be trained to directly produce energy-efficient implementations. For example, training objectives can be extended to penalize inefficient code patterns, high-complexity logic, or unnecessary computational operations, encouraging the generation of optimized and resource-efficient solutions. Green reinforcement learning approaches may also be applied, where models receive reward signals based on improvements in runtime efficiency, reduced energy consumption, or lower carbon footprint of the generated code. Moreover, parameter-efficient adaptation methods can further reduce training and deployment cost, making energy-aware code generation more practical. By embedding energy-efficiency objectives directly into model training and inference processes, future research can enable LLMs to proactively generate energy-efficient code rather than relying solely on post-hoc energy evaluation.

\subsection{Limited Focus on Small Models}

Another important observation from our review is the limited focus on small language models in energy-aware code generation research. Only two studies exclusively evaluate small models, while four additional studies include small models alongside medium and large models for comparative analysis. This indicates that the majority of existing research remains centered on large-scale models, with relatively limited dedicated investigation of lightweight architectures.

Small language models generally require less computational power and energy during inference, making them promising for energy-efficient code generation, particularly in edge and resource-constrained environments. However, their trade-offs among correctness, energy consumption, and overall efficiency remain insufficiently studied, as most existing work treats them mainly as secondary baselines. Future research should therefore conduct controlled evaluations of small models and explore techniques such as parameter-efficient adaptation, lightweight fine-tuning, quantization, and prompt optimization to improve code-generation performance while maintaining low energy consumption.

\subsection{Need for Energy-Oriented Benchmarks}
Another important observation from this study is that none of the selected works introduce a benchmark specifically designed to evaluate energy-efficiency. Instead, all studies use existing general-purpose benchmarks and add energy-related metrics, such as energy consumption or carbon footprint, as additional evaluation measures. Although dedicated efficiency-oriented benchmarks such as EffiBench have been developed to evaluate runtime and memory efficiency of generated code, dedicated benchmarks for directly evaluating the energy consumption and broader environmental impact of LLM-generated code remain limited. In particular, the reviewed literature does not provide a standardized benchmark that jointly evaluates energy consumption, carbon-related indicators, and hardware-dependent effects across different execution environments. This indicates a clear research gap in the field. In this review, we define an \textit{energy-efficiency-oriented benchmark} as a benchmark that explicitly supports the evaluation of energy-efficient code generation, for example, by providing energy-efficient reference implementations against which LLM-generated solutions can be compared, energy-aware tasks or prompts, and standardized procedures for measuring energy consumption across generated solutions. Such benchmarks could further incorporate carbon-related indicators, resource usage, and hardware characteristics. The lack of standardized energy-efficiency-oriented benchmarks limits comparison across studies and represents an important opportunity for future research.

\subsection{Energy Measurement and Cross-Platform Evaluation}

Our review shows considerable variation in how the energy consumption of LLM-generated code is measured. Most selected studies rely on software-based tools, including Python-based libraries, \textit{perf}, and Intel RAPL, while only a small number use external hardware power monitors. Software-based approaches are practical and accessible and can support relative comparisons between generated solutions when applied consistently under controlled experimental conditions. However, energy values obtained using different measurement tools, hardware configurations, and experimental settings may not be directly comparable across studies.

A related gap identified in our review is the limited evaluation across different hardware platforms. Most studies evaluate generated code on a single hardware configuration, while only one selected study considers two different platforms. Since the energy consumption of the same software can vary with processor architecture, power management, compiler settings, and other system characteristics, findings from a single platform may not generalize to other execution environments. Future studies should therefore adopt consistent and transparent measurement protocols and clearly report the tools, hardware configurations, and experimental settings used. Cross-platform evaluations across CPUs, GPUs, edge devices, and other environments would help determine whether the relative energy efficiency of LLM-generated code remains consistent across platforms, improving the comparability, reproducibility, and generalizability of findings.

\subsection{End-to-End Energy Efficiency of LLM Generated Code}
Future research should consider the end-to-end energy efficiency of LLM-based code generation by accounting for both the energy consumed during code generation and the energy consumed when the generated code is executed. Evaluating only the final program may overlook important differences in the generation process. For example, one model may produce an energy-efficient solution in a single attempt, while another may require multiple generations, prompt refinements, or candidate evaluations before reaching a comparable solution. Although both models may generate code with similar execution energy, their total energy cost can differ substantially.

Future studies should therefore jointly evaluate model inference energy and downstream code-execution energy. This would enable a more complete assessment of whether the energy spent during generation, including repeated prompting, or iterative refinement, is justified by the energy savings achieved during program execution. Such an end-to-end perspective would be particularly important for code that is executed repeatedly or deployed at scale, where downstream savings may offset the additional energy required during generation.

\subsection{Long-Term Energy Impact and Evaluation Metrics}

Another important direction is to move beyond short-term energy measurements and consider the long-term operational impact of LLM-generated code. Most studies evaluate energy consumption over individual executions or benchmark runs, whereas even small energy differences can become significant when code is executed repeatedly, deployed at scale, or runs continuously in production. Future studies should therefore examine how per-execution energy differences translate into cumulative energy savings under realistic workloads, execution frequencies, and deployment scenarios.

Our review also shows that most studies rely on conventional metrics such as energy consumption, runtime, power, and memory usage. While useful, these measures may not fully capture characteristics specific to LLM-based code generation, such as repeated prompting, multiple candidate generation, iterative optimization, and correctness filtering. Future work could therefore consider metrics such as energy per correct or accepted solution, energy savings relative to reference implementations, cumulative deployment energy, or net energy benefit. For example, a model that requires several attempts to produce an energy-efficient solution may have a higher overall energy cost than one that produces a comparable solution on the first attempt. More comprehensive evaluation should therefore account for both the energy required to obtain an efficient solution and the energy consumed during its subsequent execution, while also considering correctness, runtime, memory usage, workload scale, and deployment frequency.

\subsection{Energy-Aware Prompt Design}

Another promising direction is the systematic design of prompts that explicitly encourage LLMs to generate energy-efficient code. Existing studies show that prompting strategies can influence the efficiency of generated solutions, but there is still limited understanding of which prompt characteristics consistently lead to lower energy consumption while preserving correctness. Future research should therefore move beyond general prompt engineering and investigate prompts specifically designed around energy-related objectives.

Such prompts could instruct models to minimize unnecessary computation, reduce memory allocations, select more efficient algorithms and data structures, avoid inefficient API usage, or consider runtime and energy constraints during generation. Researchers could also compare direct energy-aware instructions with more structured approaches, such as providing examples of efficient implementations, requesting complexity analysis before generation, or using profiling and energy feedback to iteratively refine the produced code.
\section{Limitations and Threats to Validity}

\subsection{Construct Validity}

Construct validity concerns whether the key concepts of this review, particularly the energy efficiency of LLM-generated code and LLM-based code generation, have been correctly defined and applied during the study selection and analysis process. A challenge in this area is the wide range of terminology used in the literature. For instance, some studies may focus on energy efficiency, energy consumption, carbon emissions, resource usage, or green software engineering without directly using the same terminology adopted in this review. Likewise, LLM-based systems may be described using terms such as AI-assisted programming, program synthesis, or code completion. Because of these variations, there is a risk of unintentionally including studies that fall outside the intended scope or excluding relevant studies due to differences in wording.

This SLR specifically focuses on the energy efficiency of LLM-generated code rather than software sustainability in its broader sense. Software sustainability is multidimensional and may include attributes such as maintainability, evolvability, embodied environmental impacts, water consumption, and hardware lifecycle considerations. These dimensions are outside the primary scope of this review unless they are directly examined in relation to the energy or resource efficiency of generated code. Similarly, studies that investigate runtime or performance optimization without an empirical energy-related evaluation are not included, even though such improvements may indirectly contribute to lower energy consumption. This operational boundary was adopted to maintain a consistent and measurable construct across the selected studies.

To address potential construct validity issues, we applied several quality assurance steps as outlined in Section III. While developing the search string, we used a set of manually identified studies as benchmarks to ensure that the search terms properly captured the intended concepts. The search string was refined through multiple iterations to improve coverage and include representative studies. During the selection process, predefined inclusion and exclusion criteria were applied consistently. Any unclear or borderline cases were carefully reviewed and discussed to maintain a shared understanding of the study scope. Similarly, structured data extraction forms were developed and refined through iterative discussion to ensure consistent data collection and classification. These forms were used to systematically record evaluation metrics, model adaptation techniques, and energy-related characteristics. Although these steps do not eliminate the risk of missing or misinterpreting relevant studies, they help ensure a systematic and consistent identification of the literature included in this review.

\subsection{Internal Validity}
Internal validity concerns whether the conclusions of this review are properly supported by the evidence presented in the selected studies. In this work, potential threats mainly occur during the study selection and data extraction stages. These risks include selecting studies that do not fully meet the criteria, excluding relevant studies by mistake, and recording incorrect information due to unclear reporting, misunderstanding of the content, or subjective interpretation.

To minimize these risks, we followed clearly defined inclusion and exclusion criteria throughout the screening process. Any uncertain cases were carefully examined and discussed to ensure consistent decisions. We also used structured data extraction forms to systematically collect and classify information such as evaluation metrics, model size categories, prompt engineering techniques, and sustainability aspects. Only information explicitly stated in the papers was recorded. If details were not clearly reported, we did not make assumptions. For instance, when a study used GitHub Copilot without specifying the underlying LLM, it was categorized as “not specified.” Although these steps cannot completely remove all bias, they help maintain a systematic, transparent, and consistent review process. Also, two of the included primary studies were co-authored by the corresponding author of this SLR, which may introduce a potential risk of selection or interpretation bias. To mitigate this risk, these studies were evaluated using the same predefined inclusion and exclusion criteria, quality assessment, data extraction, and synthesis procedures applied to all other primary studies.

Moreover, a limitation of the snowballing process is that it was conducted by a single reviewer, and complete intermediate records of the total number of references and citing papers screened were not retained. Consequently, inter-rater agreement could not be calculated for this stage, and the full snowballing screening flow cannot be reconstructed quantitatively. To reduce this risk, the same predefined inclusion and exclusion criteria used in the main study-selection process were applied during snowballing. Snowballing was also used as a complementary search strategy to identify potentially missed studies rather than as the primary identification mechanism.

\subsection{External Validity}

External validity concerns whether the selected studies represent the existing body of knowledge on the energy efficiency of LLM-generated code. In this review, we included peer-reviewed publications as well as relevant preprints from arXiv, given that this research area is relatively new and rapidly evolving. However, we excluded non-English publications, unpublished student theses, technical blogs, and other forms of grey literature not indexed in the selected databases. As a result, some relevant studies may not have been captured.

The database search was conducted in October 2025. Therefore, studies published, indexed, or made publicly available after this search period were not considered in the review. It is also possible that some studies had been accepted for publication but were not yet publicly accessible or indexed at the time of the search. Consequently, our findings reflect the state of the literature available through the selected databases as of October 2025.

Moreover, the final study set includes six preprints. Although preprints were included to capture recent empirical work in this rapidly evolving area, they have not necessarily undergone formal peer review and may therefore introduce additional uncertainty regarding study quality. To reduce this risk, the same inclusion and exclusion criteria, quality assessment, data extraction, and synthesis procedures were applied to both peer-reviewed studies and preprints. Despite these limitations, the search strategy, inclusion and exclusion criteria, study-selection procedure, and data extraction forms presented in this study provide a transparent framework that can be reused or extended by future researchers to replicate or update the review.

\section{Conclusions}
LLMs are increasingly being integrated into software engineering workflows to assist developers with tasks such as code generation, completion, translation, and bug fixing. While significant research has focused on improving the functional correctness and productivity gains of these models, comparatively little attention has been paid to the energy-efficiency of the code they produce. We conduct an SLR to examine whether LLMs can generate energy-efficient code. We follow a structured process that includes comprehensive database searches and a snowballing method, and clear inclusion and exclusion criteria are applied across multiple screening stages. For the selected studies, we extract data using predefined forms to ensure consistency and accuracy. Our findings show that research in this area is still limited. Most studies focus on LLMs, while small models receive little attention. Energy evaluation is mainly performed at the software level, and only a few studies use external hardware tools for direct physical measurement. In addition, research is largely limited to a small number of programming languages and general-purpose applications. There are no dedicated benchmarks specifically designed to measure the energy-efficiency of LLM-generated code, and energy-aware fine-tuning remains an open area for future work. To advance this field, future research should develop dedicated energy related benchmarks, adopt more accurate energy measurement techniques, and explore energy-aware training and fine-tuning strategies. Addressing these gaps is essential to ensure that the growing use of LLMs in software development aligns with long-term environmental sustainability goals.

\bibliographystyle{IEEEtran}
\bibliography{references}

\end{document}